\documentclass[twocolumn,showpacs,amsbsy,amstex,amssymb,prb]{revtex4}
\usepackage{graphicx}
\usepackage{epstopdf}

\newcommand{\dd}{{\bf d}}
\newcommand{\br}{{\bf r}}
\begin{document}
\title{Mapping the braiding properties of the Moore-Read state}
\author{Emil Prodan$^1$ and Frederick Duncan M. Haldane$^2$}
\address{$^1$Department of Physics, Yeshiva University, New York, NY 10016} 
\email{prodan@yu.edu}
\address{$^2$Department of Physics, Princeton University, Princeton, NJ 08544}

\begin{abstract}

In this paper we explore the braiding properties of the Moore-Read fractional Hall sequence, which amounts to computing the adiabatic evolution of the Hall liquid when the anyons are moved along various trajectories. In this work, the anyons are pinned to precise spatial configurations by using specific external potentials.  Such external potentials break the translational symmetry and it appears that one will be forced to simulate the braidings on the entire many-body Hilbert space, an absolutely prohibitive scenario. We demonstrate how to overcome this difficulty and obtain the exact braidings for fairly large Hall systems. For this, we show that the incompressible state of a general $(k,m)$ fractional Hall sequence can be viewed as the unique zero mode of a specific Hamiltonian $H^{(k,m)}$, whose form is explicitly derived by using k-particles creation operators. The compressible Hall states corresponding to $n$$\times$$k$ anyons fixed at $w_1$,\ldots,$w_{nk}$ are shown to be the zero modes of a pinning Hamiltonian $H^{(k,m)} _{w_1,\ldots,w_{nk}}$, which is also explicitly derived. The zero modes of $H^{(k,m)} _{w_1,\ldots,w_{nk}}$ are shown to be contained in the space of the zero modes of $H^{(k,m)}$. Therefore, the computation of the braidings can be done entirely within this space, which we map out for a number of Hall systems. Using this efficient computational method, we study various properties of the Moore-Read states. In particular, we give direct confirmation of their topological and non-abelian properties that were previously implied from the underlying Conformal Field Theory (CFT) structure of the Moore-Read state. 
\end{abstract}


\maketitle

\section{Introduction}

The recently discovered Fractional Hall states at filling factors $\nu = 5/2$,\cite{Willett:1987fu} and $\nu$=12/5,\cite{Xia:2004gb} are thought to contain anyons obeying Non-Abelian statistics. It was argued\cite{Rezayi:lh} that these states are part of the Moore-Read sequence,\cite{Moore:1991cr} and of the Read-Rezayi sequence at $k$=3,\cite{Read:1999oq} respectively. Both states received sustained attention from the theoretical and experimental condensed matter community because they can provide the key to the physical realization of a Topological Quantum Computer (TQC). For this reason, there is a concentrated effort on both theoretical and experimental fronts for finding direct confirmations of the Non-Abelian and Topological properties of these Fractional Hall states.\cite{C.-Chamon:1997bh,Isakov:1999dq,Safi:2001cr,Vishveshwara:2003nx,Kane:2003kl,Kim:2005oq,Law:2006tg,Bonderson:2006qf,Feldman:2006ve}

In spite of an overwhelming indirect evidence, the non-Abelian statistics for the Moore-Read sequence was not considered resolved\cite{Read:2007cq} until the relatively recent numerical confirmation of Ref.~\onlinecite{Tserkovnyak:2003vn}, where Monte-Carlo techniques were used to compute a small number of adiabatic mononodromies. The present paper reports the results of an exact diagonalization study, which allows us to take a direct and unprecedented look into the non-Abelian and topological properties of  the Moore-Read sequence. We have results for systems containing 2 and 4 anyons. A model Hamiltonian (reported here for the first time) is used to pin down the anyons at specified locations and to move them adiabatically along different paths. Among our results, the reader will find the following:

- a map of the Abelian and non-Abelian adiabatic curvature experienced by an itinerant anyon when the rest of the anyons are kept at fixed locations.

- a map of a newly introduced Twist Density, which measures the twist of the zero modes space during the adiabatic braidings.

- a direct proof that the adiabatic curvature is strongly localized near the fixed anyons, thus confirming the topological properties of adiabatic braids.

- a direct proof that the monodromy corresponding to a braiding in which one anyon loops around another anyon is in perfect agreement with the CFT prediction.

- a direct proof that the zero modes space splits according to the fusion rules of the underling Conformal Field Theory structure of the Moore-Read state.

- a direct proof that the conformal blocks can be distinguished by bringing two anyons together and by measuring the electron density for the fused anyons.

Let us briefly mention the relevance of our study for for TQC. The logic gates in TQC algorithms can be generated by sequences of braids. The specific sequences of braids can be computed using the Solovay-Kitaev algorithm,\cite{Kitaev:1999tw} once a unitary representation of the anyons' braid group is provided. For the Fibonacci anyons, for example, the braiding sequences implementing the elementary logic gates were explicitely calculated in Refs.~\onlinecite{Bonesteel:2005zr} and \onlinecite{Hormozi:2007ve}. The unitary representations of the braid group can be classified by various methods. Particularly, the Ref.~\onlinecite{Slingerland:2001il} describes a general method for generating unitary representations of the braid group using quantum groups. These representations were connected to the braidings of the anyons in Fractional Hall Liquids via the argument that braiding can be computed from an effective picture in which the anyons carry an internal quantum symmetry. This symmetry can be derived from the underlying CFT structure of the Fractional Hall sequence. For example, the braiding of the Fibonacci anyons in the $\widehat{sl(2)}_k$ WZW theory can be explicitly generated using the $U_q(sl(2))$ quantum group at the root of unity $q$=$e^{2\pi i/(k+2)}$.\cite{Slingerland:2001il} It was also argued\cite{Slingerland:2001il} that the braiding of the Read-Rezayi parafermions can be explicitly generated using the $U_{q_1}(sl(2))\otimes \mathbb{CZ}_{4k,q2}$ quantum group at the roots of unity $q_1$=$e^{2\pi i/(k+2)}$ and $q_2$=$e^{-i\pi/2k}$. For the Moore-Read sequence ($k$=2), the braiding reduces to the spinor representation of the mirror reflections in the group SO($2n$), as discovered long time ago by Nayak and Wilczek.\cite{Nayak:1996mb}  

In all these algebraic derivations the anyons are assumed infinitely far apart from each other. In practice this will not be the case, and this is the main reason why the algebraic results need a direct confirmation showing that the braiding properties remain unchanged when finite anyon densities are considered. So far, our numerical results confirm that this is the case.

\section{Fractional Hall sequences: General considerations.}

The Hall sequences can be label by two integers: $k$=1,2,\ldots, and $m$=0,1,\ldots. The filling factor and the conformal charge for each sequence are given by:
\begin{equation}\label{nu}
\nu = \frac{k}{km+2}, \ \ c = 1+ \frac{2(k-1)}{k+2}.
\end{equation}
The $k$=1, $\nu$=$1/(m+2)$ sequence corresponds to the Laughlin state,\cite{Laughlin:1983kl} which has fractionally charged anyons carrying integer flux. Their braiding is Abelian. The $k$$\ge$2 sequences have fractionally charged anyons carrying fractional flux. Their braiding is believed to be Non-Abelian. The case $k$=2 corresponds to the Moore-Read sequence;\cite{Moore:1991cr} $k$=3 and greater correspond to the Read-Rezayi sequences.\cite{Read:1999oq} Excepting $k$=1, 2 and 4, all sequences support universal quantum computation. 

We define the $k$-particle creation operators, $\eta_M^{(k,m)\dagger}$, generating (from the vacuum) $k$-particles in a Laughlin state (we set the magnetic length to 1):
\begin{equation}\label{kcreation}
\begin{array}{c}
\langle {\bf r}_1, \ldots, {\bf r}_k |\eta_M^{(k,m)\dagger}|0\rangle = (z_1+\ldots + z_k)^M  \medskip \\
\times \prod \limits _{i<j \le k} (z_i-z_j)^m e^{-\frac{1}{4}\sum _{i=1}^k |z_i|^2},
\end{array}
\end{equation}
where $z$=$x$+$iy$ is the complex representation of the position $\br$=$(x,y)$ in the 2-dimensional plane, and $M$ and $m$ are integers larger or equal to zero. The Fractional Hall sequences can be described as follows.

{\it Incompressible states.} We claim that the incompressible Hall state for an arbitrary ($k$,$m$), originally defined in terms of certain correlators of the $\mathbb{Z}_k$  parafermion conformal algebra, is the highest electron density state satisfying
\begin{equation}\label{kbc}
\begin{array}{l}
\eta_M^{(2,m')} |\Psi^{(k,m)}\rangle = 0, \ \forall \ M \ge 0 \ \text{and} \ m'<m, \medskip \\
\eta_M^{(k+1,m)} |\Psi^{(k,m)}\rangle = 0, \ \ \forall \ M \ge 0.
\end{array}
\end{equation}
These incompressible states occur at electron densities given in Eq.~(\ref{nu}).

Let us elaborate on the above conditions. From the definition given in Eq.~\ref{kcreation}, we have
\begin{equation}\label{st1}
\begin{array}{l}
[\eta_M^{(2,m')} \Psi^{(k,m)}](z_3,\ldots,z_N) = \int d^2 z_1 \int d^2 z_2\medskip\\
\times  (z_1^*+z_2^*)^M
(z_1^*-z_2^*)^{m'}e^{-\frac{1}{4}(|z_1|^2+|z_2|^2)} \medskip\\
\times \Psi^{(k,m)}(z_1,z_2,\ldots,z_N) .
\end{array}
\end{equation}
The general structure of $\Psi^{(k,m)}$ is
\begin{equation}
\begin{array}{l}
\Psi^{(k,m)}(z_1,\ldots,z_N) = \tilde{\Psi}_k(z_1, \ldots ,z_N) \medskip \\
\times \prod \limits _{i<j \le N} (z_i-z_j)^m e^{-\frac{1}{4}\sum _{i=1}^N |z_i|^2},
\end{array}
\end{equation}
where $\tilde{\Psi}_k(z_1, \ldots ,z_N)$ is the correlation function of $N$ number of ${\bf Z}_k$ parafermion fields (we include in $\tilde{\Psi}_k$ the factors that make the correlation function non-singular). We shall see that the first condition of Eq.~(\ref{kbc}) relates to the part of the wavefunction contained in the second row of the above equation. Indeed, the right hand side of Eq.~(\ref{st1}) reduces to
\begin{equation}
\begin{array}{l}
\prod \limits _{2<i<j } (z_i-z_j)^m e^{-\frac{1}{4}\sum _{i=3}^N |z_i|^2} 
 \int d^2 z_1 \int d^2 z_2 \medskip \\ 
\times  (z_1^*+z_2^*)^M |z_1-z_2|^{2m'}e^{-\frac{1}{2}(|z_1|^2+|z_2|^2)} \medskip \\
\times (z_1-z_2)^{m-m'} \prod \limits _{i = 3}^N (z_i-z_1)^m(z_i-z_2)^m \medskip \\
\times \tilde{\Psi}_k(z_1, \ldots ,z_N)
\end{array}
\end{equation}
The multiple integral projects the part of the integrand that is invariant  to 4-dimensional rotations in the (${\bf r}_1$,${\bf r}_2$) sub-space and we argue that this part is zero. We focus to rotations around the axis connecting ${\bf r}_1+{\bf r}_2$ to the origin. The second row of the above equation is invariant to such rotations but the invariant part of the remaining rows is identically zero. Indeed, the third and fourth rows are analytic functions in all complex variables, hence their product admit an expansion like $\sum_\alpha c_\alpha (z_1-z_2)^\alpha$, where $c_\alpha$ may depend on $z_1+z_2$, $z_3$, \ldots,$z_N$. The invariant part of such expansion is equal to $c_0$, but $c_0$ is identically zero when $m'$$<$$m$. This proves the first condition of Eq.~\ref{kbc}.

We consider now the second condition, which says that
\begin{equation}\label{st2}
\begin{array}{l}
[\eta_M^{(k+1,m)} \Psi^{(k,m)}](z_{k+2},\ldots,z_N) = \int d^2 z_1 \ldots \int d^2 z_{k+1}\medskip\\
\times  (z_1^*+\ldots,+z_{k+1}^*)^M
\prod \limits _{i<j \le k+1} (z_i^*-z_j^*)^m e^{-\frac{1}{4}\sum \limits_{i=1}^{k+1} |z_i|^2} \medskip\\
\times \Psi^{(k,m)}(z_1, \ldots ,z_N).
\end{array}
\end{equation}
is identically zero. The right hand side of Eq.~(\ref{st2}) is equal to:
\begin{equation}\label{preliminary}
\begin{array}{l}
\prod \limits _{k+2 \le i<j } (z_i-z_j)^m e^{-\frac{1}{4}\sum \limits _{i>k+1} |z_i|^2}
\int d^2 z_1 \ldots \int d^2 z_{k+1}\medskip\\
\times  (z_1^*+\ldots,+z_{k+1}^*)^M
\prod \limits _{i<j \le k+1} |z_i-z_j|^{2m} e^{-\frac{1}{2}\sum \limits _{i=1}^{k+1} |z_i|^2} \medskip\\
\times \tilde{\Psi}_k(z_1, \ldots ,z_N)\prod \limits _{i = 1}^{k+1}\prod \limits _{j > k+1} (z_i-z_j)^m.
\end{array}
\end{equation}
The third row is an analytic function in all complex arguments, hence admits an expansion such as:
\begin{equation}
\sum_\alpha c_\alpha \prod\limits_{i<j \le k+1} (z_i-z_j)^{\alpha_{ij}},
\end{equation}
where the coefficients may depend on $z_1+ \ldots +z_{k+1}$, $z_{k+2}$, \ldots,$z_N$. The index $\alpha$ refers to the collection of indices $\alpha_{ij}$ appearing inside the product. The only nonzero contribution to the integral of Eq.~\ref{preliminary} comes from the term $\alpha=0$, as one can see by switching to the coordinates $s=(z_1+\ldots +z_{k+1})/(k+1), \tilde{z}_i=z_i-s$. But there is no such term because $\tilde{\Psi}_k(z_1, \ldots ,z_N)$ cancels identically when $k$$+$$1$ particles come to the same point in space.\cite{Read:1999oq}

We reformulate Eq.~\ref{kbc} in terms of the creation operators $\eta^{(k,m)}(w)^\dagger$ generating $k$-particle Laughlin clusters centered at different $w$'s:
\begin{equation}
\begin{array}{l}
\langle {\bf r}_1, \ldots, {\bf r}_k |\eta^{(k,m)}(w)^\dagger|0\rangle = \medskip \\
\prod \limits _{i<j \le k} (z_i-z_j)^m e^{-\frac{1}{4}\sum \limits _{i=1}^k |z_i-w|^2}.
\end{array}
\end{equation}
The incompressible states are then uniquely defined as the highest electron density states satisfying
 \begin{equation}
 \begin{array}{l}
\eta^{(2,m')}(w) |\Psi^{(k,m)}\rangle = 0, \ \ \forall \ w \ \text{and} \ m'<m, \medskip \\
\eta^{(k+1,m)}(w) |\Psi^{(k,m)}\rangle = 0, \ \ \forall \ w.
\end{array}
\end{equation}

{\it Compressible states.} The lower electron density states satisfy the same conditions but the uniqueness is, of course, lost. A lower density state $\Psi^{(k,m)}_{w_1\ldots w_{nk}}$ with $n\times k$ anyons located at $\{w\}$=$w_1$,\ldots,$w_{nk}$ has the general form
\begin{equation}\label{qh}
\begin{array}{l}
\Psi^{(k,m)}_{\{w\}}(z_1,\ldots,z_N)= \tilde{\Psi}_{\{w\}}(z_1, \ldots ,z_N) \medskip \\
\times \prod \limits _{i<j \le N} (z_i-z_j)^m e^{-\frac{1}{4}\sum \limits _{i=1}^N |z_i|^2},
\end{array}
\end{equation}
where $\tilde{\Psi}_{\{w\}}(z_1,\ldots,z_N)$ is a certain correlator.\cite{Read:1999oq} This correlator is a multi-valued function in $w$'s. Each branch defines a different state. All these states have the  anyons fixed at the same locations $\{w\}$. The degeneracy of the states with fixed anyon positions is discussed in details in Ref.~\onlinecite{Slingerland:2001il}, for arbitrary $k$. The correlators $\tilde{\Psi}_{\{w\}}(z_1,\ldots,z_N)$ have the property of vanishing whenever $k$ particles meet at any of the $w$'s.\cite{Read:1999oq} 

We now show that these states can be defined by an additional condition to Eq.~(\ref{kbc}), namely:
\begin{equation}\label{any}
\eta^{(k,m)}(w_\mu) |\Psi^{(k,m)}_{\{w\}}\rangle = 0, \ \ \forall \ \mu = 1,\ldots,nk.
\end{equation}
Indeed, we have
\begin{equation}
\begin{array}{l}
[\eta^{(k,m)}(w_\mu) \Psi^{(k,m)}_{\{w\}}](z_{k+1},\ldots,z_N) = \medskip \\
 \int d^2 z_1 \ldots \int d^2 z_k 
\prod \limits _{i<j \le k} (z_i^*-z_j^*)^m e^{-\frac{1}{4}\sum \limits_{i=1}^{k} |z_i-w_\mu|^2} \medskip \\
\times \tilde{\Psi}_{\{w\}}(z_1, \ldots ,z_N)\prod \limits _{i<j \le N} (z_i-z_j)^m e^{-\frac{1}{4}\sum \limits _{i=1}^N |z_i|^2}.
\end{array}
\end{equation}
The right hand side can be written as:
\begin{equation}
\begin{array}{l}
\prod \limits _{k<i<j} (z_i-z_j)^m e^{-\frac{1}{4}\sum \limits _{i>k} |z_i|^2}
\int d^2 z_1 \ldots \int d^2 z_k \medskip \\
\times \prod \limits _{i<j \le k} |z_i-z_j|^m e^{-\frac{1}{4}\sum \limits_{i=1}^{k} |z_i-w_\mu|^2+|z_i|^2} \medskip \\
\times \tilde{\Psi}_{\{w\}}(z_1, \ldots ,z_N)\prod \limits _{i=1}^k\prod \limits _{j>k} (z_i-z_j)^m.
\end{array}
\end{equation}
The integral can be show to be identically zero by using the clustering properties of the correlator.

Of course, we also need to prove that Eqs.~(\ref{kbc}) and (\ref{any}) are not satisfied by states other than the quasi-hole states of Eq.~(\ref{qh}). For $k$=2 we have verified numerically (in the sphere geometry) that the number of linearly independent states satisfying these conditions is precisely equal to $2^{n-1}$, as it should.\cite{Read:1996nx} A similar study was conducted for the $k=3$ Hall sequence.\cite{Hui:hc}

\begin{figure}
  \includegraphics[width=8.6cm]{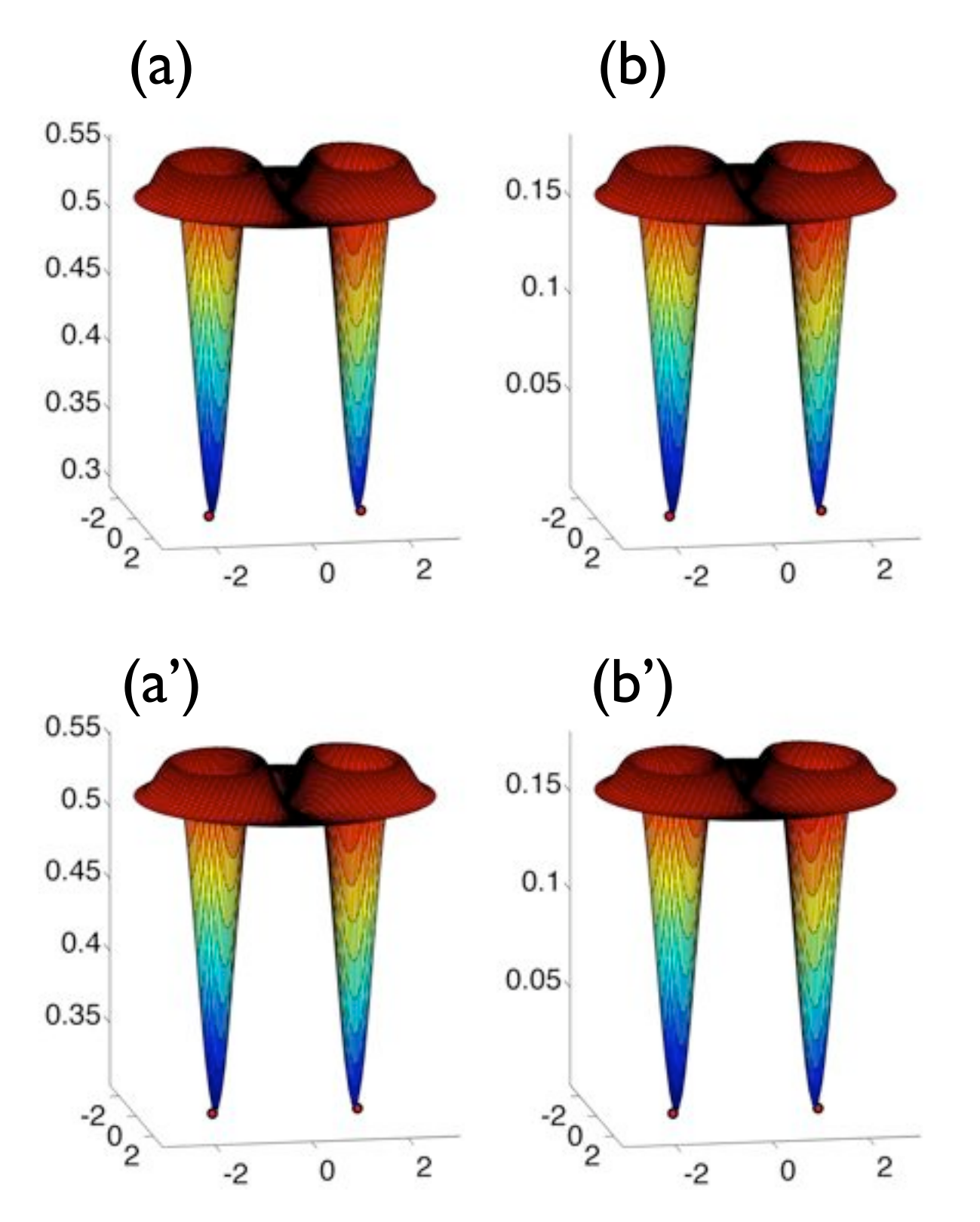}\\
  \caption{(Color online) a) and a') The particle density and b) and b') the pair amplitude corresponding to the unique zero mode of the pinning Hamiltonian $H(w_1,w_2)$$\downharpoonright$$_{{\cal H}_0}$. The first row corresponds to the odd or S=1 sector  ($N$=15, $N_\phi$=29) and the second row corresponds to the even or S=0 sector ($N$=16, $N_\phi$=31). The position of the probes is indicated by red dots.}\label{example1}
\end{figure}

\section{The pinning Hamiltonian} 

Based on the description given in the previous Section, we can generate simple model Hamiltonians for each Fractional Hall sequence. Such a task was previously undertaken in Ref.~\onlinecite{Simon:2007kl}.

We claim that all incompressible fractional Hall states corresponding to given $k$ and $m$ are zero energy states for the following Hamiltonian:
\begin{equation}\label{rm}
\begin{array}{l}
H^{(k,m)} = \lambda \int d^2 w \ \eta^{(k+1,m)}(w)^\dagger \eta^{(k+1,m)}(w) \medskip \\
+\sum\limits _{m'<m} \lambda_{m'} \int d^2w \ \eta^{(2,m')}(w)^\dagger \eta^{(2,m')}(w),
\end{array}
\end{equation}
where $\lambda$'s must be all positive. For $k$=2 we have also verified that the opposite is true, namely that the zero modes space of the above Hamiltonian contains all the fractional Hall states and that its dimension coincide with the theoretical value derived in Ref.~\onlinecite{Read:1996nx}. A similar study exists for $k$=3 sequence.\cite{Hui:hc}

We also claim that a lower density state with $n$$\times$$k$ anyons present at $w_1$,\ldots,$w_{nk}$ is a zero energy state of the following pinning Hamiltonian:
\begin{equation}
\begin{array}{l}
H^{(k,m)} _{w_1,\ldots,w_{nk}}= \lambda \int d^2 w \ \eta^{(k+1,m)}(w)^\dagger \eta^{(k+1,m)}(w) \medskip \\
+\sum\limits _{m'<m} \lambda_{m'} \int d^2w \ \eta^{(2,m')}(w)^\dagger \eta^{(2,m')}(w) \medskip \\
+ \sum\limits_{\mu = 1}^{nk} \lambda_\mu \ \eta^{(k,m)}(w_\mu)^\dagger \eta^{(k,m)}(w_\mu),
\end{array}
\end{equation}
where, again, all $\lambda$'s are considered positive. It is useful to think of the pinning Hamiltonian $H(\{w\})$ as simulating a set of $n$$\times$$k$ external probes that have been placed at the locations $w_1,\ldots,w_{nk}$. Again, for $k$=2 and small number of units of flux added to the fundamental value of the magnetic flux, we have verified numerically (on the sphere) that the dimension of the zero energy modes space of the above Hamiltonian is precisely equal to $2^{n-1}$. A similar study exists for the $k$=3 sequence.\cite{Hui:hc}

Given the particular form of the pinning Hamiltonian, namely, the fact that each term is positive definite, we have the following crucial observation: the zero energy states of the pinning Hamiltonian, for arbitrary anyon configuration $\{w\}=w_1$,\ldots,$w_{nk}$, {\it can be computed in two steps without involving any approximation.} Here is how: 

\begin{enumerate}

\item Construct the null space ${\cal H}_0$ of the Hamiltonian $H^{(k,m)}$ given in Eq.~(\ref{rm}).

\item Restrict the pinning Hamiltonian:
\begin{equation}\label{pinning}
H(\{w\}) =\sum\limits_{\mu = 1}^{nk} \eta^{(k,m)}(w_\mu)^\dagger \eta^{(k,m)}(w_\mu)
\end{equation}
to ${\cal H}_0$ and construct its null space ${\cal H}_0(\{w\})$, which contains all the Hall states with anyons pinned at $w_1,\ldots,w_{nk}$.
\end{enumerate}

As we shall exemplify later, the dimension of the many-body Hilbert space increases extremely fast with the number of electrons. For this reason, any attempt of direct diagonalization of the full pinning Hamiltonian on the full Hilbert space is futile. Now, the difference between the pinning Hamiltonian and $H^{(k,m)}$ is that the latest is translational invariant. In the sphere geometry,\cite{Haldane:1983fk} this means that $H^{(k,m)}$ commutes with the total angular momentum ${\bf L}$. Thus, to generate the null space ${\cal H}_0$ in the sphere geometry, we need to search only for the zero modes of highest weight, i.e. the ones in the null space of $L^+$ operator. Once  this step is completed, we can generated the full ${\cal H}_0$ by successively  applying $L^-$ operator on the highest weight zero modes. This program was numerically implemented using standard techniques.

\begin{figure}
  \includegraphics[width=8cm]{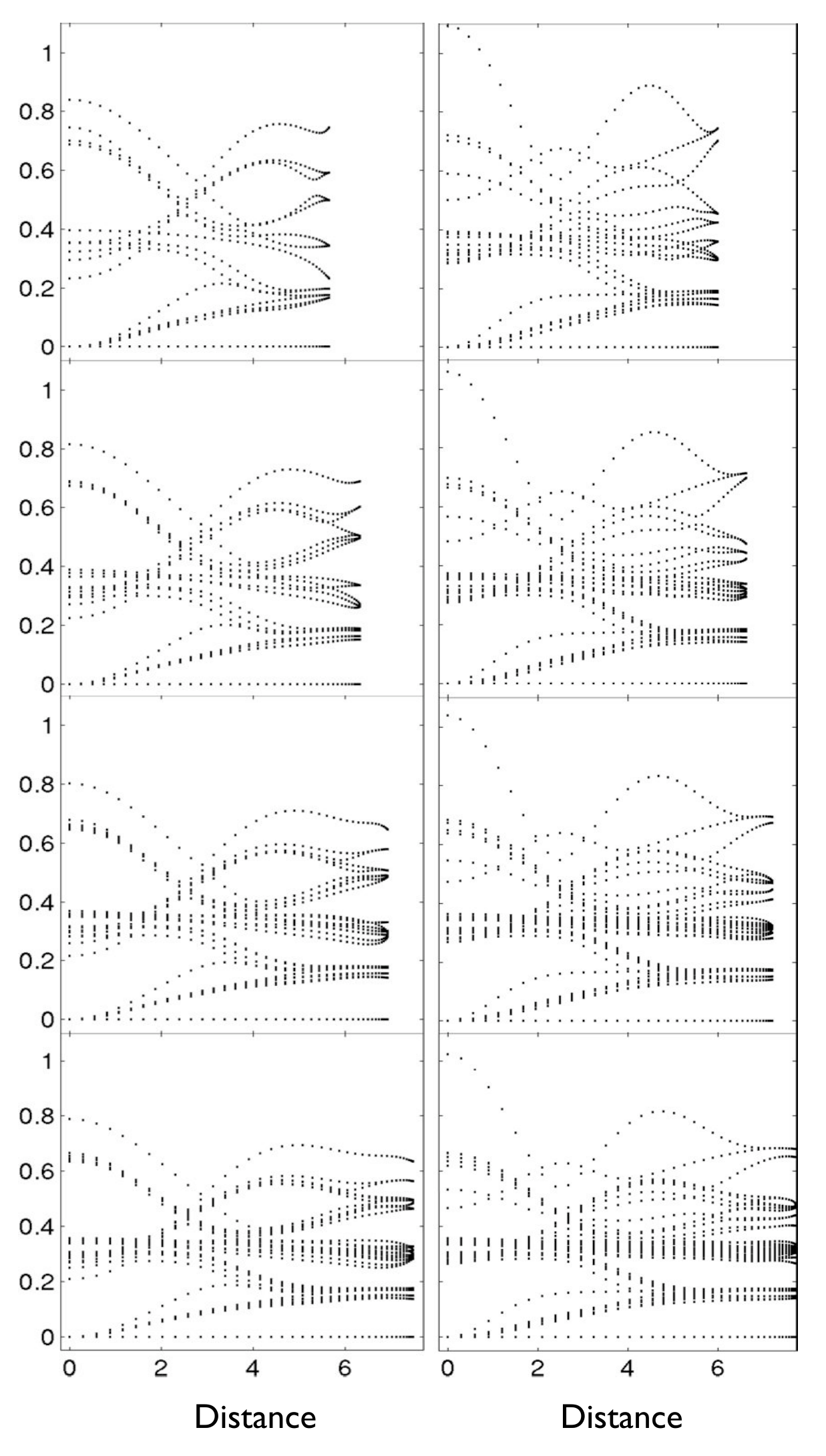}\\
  \caption{(Color online) The spectrum of $H(w_1,w_2)$$\downharpoonright$$_{{\cal H}_0}$ as function of distance between the probes, which are moved along the meridians $(\theta,\phi=0)$ and $(\theta,\phi=\pi)$, with $\theta$ increasing from 0 to $\pi/2$. The left /right column corresponds to odd/even number of electrons. Starting from the top, the left panels correspond to $N/N_\phi$=9/16, 11/20, 13/24, 15/28 and the right panels to $N/N_\phi$=10/18, 12/22, 14/26, 16/30.}
 \label{SpectrumVdistance}
\end{figure}

\section{Results for the Moore-Read sequence}

Here we show a set of results for the Moore-Read sequence $k$=2 and $m$=1 in the first Landau level. The calculations were performed with the sphere geometry, where we can work with a finite number of electrons. The number of available orbitals in the first Landau level is $N_{\text{orb}}$=$N_\phi$+1, where $N_\phi$ is the magnetic flux (in the quantum units of flux) passing through the surface of the sphere. We have to separately discuss the case of odd and even number of electrons. The incompressible state exists only for even number $N$ of electrons, and occurs when the number of flux is relation $N_\phi^0$=$2N$$-$3 with the number $N$ of electrons. For this case, $\dim({\cal H}_0)$=1.

\subsection{One pair of anyons.} When the number of flux becomes $N_\phi$=$N_\phi^0$+1, a pair of anyons is generated and
\begin{equation}\label{nr1}
\dim({\cal H}_0) = \frac{1}{2}([N/2]+1)([N/2]+2),
\end{equation}
where the square brackets indicates the integer part.\cite{Read:1996nx} We have computed and saved the ${\cal H}_0$ spaces for $N/N_\phi$=4/6, 5/8, 6/10, 7/12, 8/14, 9/16, 10/18, 11/20, 12/22, 13/24, 14/26, 15/28 and 16/30. The largest system in this sequence, the 16 electrons on 31 orbitals, has $\dim({\cal H}_0)$=45. 

If we fix the positions of the two quasi-holes at arbitrary locations $w_1$ and $w_2$, by adding the pinning potential described in Eq.~\ref{pinning}, all the states in ${\cal H}_0$ are pushed up in energy, except one state whose energy remains exactly zero. This is precisely the state $\Psi^{(k,m)}_{\{w_1,w_2\}}$ discussed above. To exemplify, we consider the largest systems we computed so far, namely, the system with 15 electrons on  29 orbitals (odd number of electrons) and the system with 16 electrons on  30 orbitals (even number of electrons). In this cases, the total many-body Hilbert spaces have staggering dimensions of 77,558,760 and 300,540,195, respectively. In this extremely large Hilbert spaces, we find a number of zero modes for $H^{(2,2)}$ equal to 36 in the first case and 45 in the second case (in total agreement with Eq.~(\ref{nr1})). If we fix $w_1$ and $w_2$ and diagonalize $H(w_1,w_2)$$\downharpoonright$$_{{\cal H}_0}$, we find one zero mode for both cases. Finding these zero modes would have been impossible without taking full advantage of the translational symmetry at the first step, when ${\cal H}_0$ was resolved.

To visualize a state, we compute the corresponding particle density and pair amplitude as functions of position on the sphere. The latest is given by the expectation value of $\eta^{(2,2)}(w)^\dagger \eta^{(2,2)}(w)$ on the zero mode. A plot of these quantities for the zero modes discussed above, is shown in Fig.~\ref{example1}. The positions of the probes were chosen as ($\theta$=$\pi/2$,$\phi$=0) and ($\theta$=$\pi/2$,$\phi=\pi$), so that we have maximum possible separation between the trapped anyons. Since the anyons are far apart, there is no visible difference between even and odd cases. Refering to the Bratelli diagram for the Moore-Read sequence,\cite{Bonderson:2006qf} the even and odd number of electrons correspond to the q-spin $S$=0 and $S$=1 sectors, respectively. Thus, we can see that there is no difference in the local properties of the wavefunctions belonging to different conformal blocks, which is precisely what one should see for a topological degeneracy. As we shall see, things look completely different when we bring the anyons close to each other. Other things to notice about Fig.~\ref{example1} are the fact that the density is finite while the pair amplitude is exactly zero at the probe locations and the fact that the two anyons appear to be totally separated.

Let us take a few lines here and explain our plots. Quantities that depend on the position on the sphere will be shown as surface plots, with the quantity of interest on the z axis. The cartesian coordinates x and y describe points of the sphere. If $\theta$ and $\phi$ are the usual angles on the sphere, then the relation between ($\theta$,$\phi$) and $(x,y)$ is given by $\theta$=$\sqrt{x^2+y^2}$ and $\phi$=$\arctan(y/x)$.

Next, let us take a look at the energy spectrum of $H(w_1,w_2)$$\downharpoonright$$_{{\cal H}_0}$ as function of probe separation, $d(\theta) = \sqrt{ N_\phi} \sin \frac{\theta}{2}$, while gradually increasing the number of electrons from 9 to 16. For each size, the probes were moved along the meridians ($\theta$,$\phi$=0) and ($\theta$,$\phi$=$\pi$), with $\theta$ increasing from 0 to $\pi/2$.  The strength of the probe potential was fixed at $\lambda$=1. The results are shown in Fig.~\ref{SpectrumVdistance}, where each panel displays a number of bands (equal to $\\dim ({\cal H}_0)$), representing the flow of the eigenvalues with the distance $d$. There is one and only one eigenvalue that remains strictly zero (within a numerical error that is less than $10^{-12}$!). The energy gap separating this zero mode from the rest of the spectrum goes to zero as the probes come closer to each other and converge to a well defined limit when the probes are moved far apart from each other. Compared to the other level spacings in the graph, the energy gap appears large. Another thing to notice is that there is a difference in the eigenvalues flow patterns when the plots for odd and even number of electrons are compared. It is also interesting to remark that the flow patterns remains almost unchanged as the size of the system is increased.

\begin{figure}
  \includegraphics[width=8.6cm]{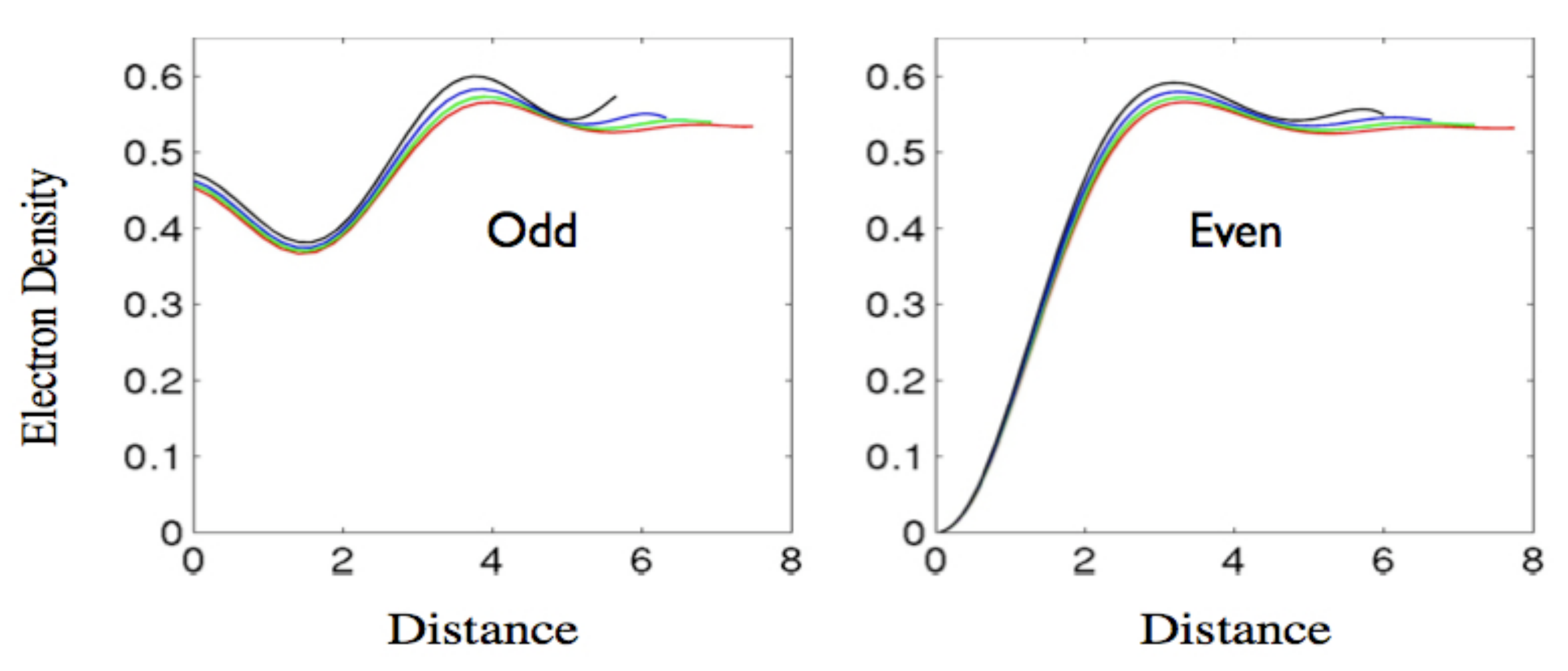}\\
  \caption{(Color online) The particle density for fused probes as function of distance from the fusing point. Left/right panel refers to odd/even number of electrons. On the left, the different curves correspond to $N/N_\phi$=9/16, 11/20, 13/24, 15/28 and, on the right, the different curves correspond to $N/N_\phi$=10/18, 12/22, 14/26, 16/30.}\label{FusedProbes}
\end{figure}

{\it Fussing the anyons.} As one can clearly see in Fig.~\ref{SpectrumVdistance}, when the anyons are fused, the lowest energy level becomes degenerate.  The electron density is not well defined, but we can still study its limit as the distance between the anyons go to zero. In this limit, the electron density becomes radially symmetric and we can plot the density as a function of the distance from the position of the fused anyons. The results are shown in Fig.~\ref{FusedProbes} for different sizes of the Hall system. The graphs reveal that the particle density is different for odd and even number of electrons. This means that one should be able to tell when a pair of anyons is in the S=0 or S=1 sector by simply fusing the anyons together and measuring the electron density.  Fig.~\ref{FusedProbes} also reveals that the electron density is rapidly converging with the size of the system.

\subsection{Two pairs of anyons}

When the number of flux becomes $N_\phi$=$N_\phi^0$+2, two pairs of anyons are created and\cite{Read:1996nx}
\begin{equation}\label{nr2}
\dim({\cal H}_0) =\frac{1}{192} \left \{
\begin{array}{l}
(N+1)(N+3)(N+5)(N+7) \\
(N+2)(N+4)^2(N+6), 
\end{array} \right . 
\end{equation}
where the first row refers to odd and the second row to even number of electrons. We have computed and saved the ${\cal H}_0$ spaces for $N/N_\phi$=4/7, 5/9, 6/11, 7/13, 8/15, 9/17, 10/19, 11/21, 12/23, 13/25, 14/27, 15/29 (16/31 resisted to us so far). The largest system in this sequence, the 15 electrons on 30 orbitals, has $\dim({\cal H}_0)$=660.

\begin{figure}
  \includegraphics[width=8.0cm]{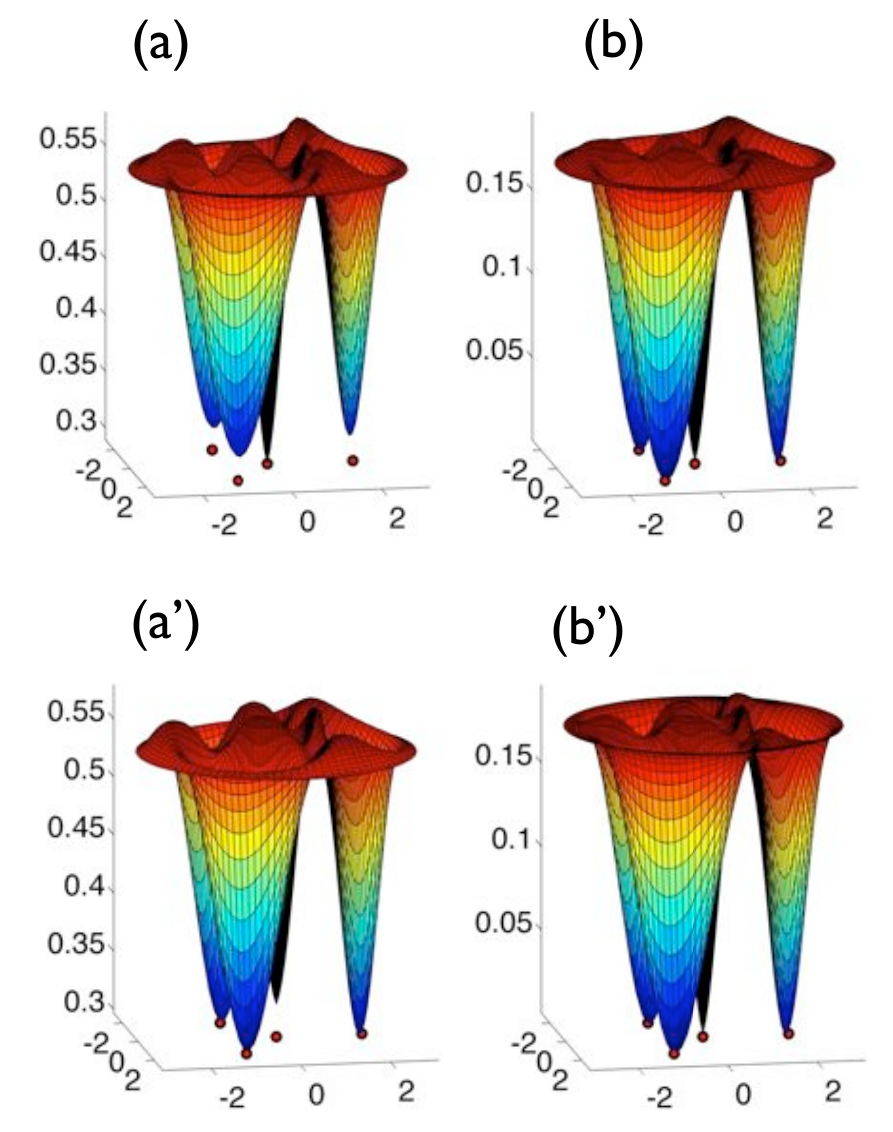}\\
  \caption{(Color online) Visualization of the two zero modes for $N/N_\phi$=15/29.  (a) and (a') shows the electron density and (b) and (b') shows the pair amplitude of the two zero modes. The zero modes were obtained by splitting the two dimensional space. The probes were fixed in a tetrahedral configuration. }
  \label{example2}
\end{figure}

It is instructive to consider an example. We pick the largest system we could compute so far, the $N$=15 and $N_{\text{orb}}$=30 case, where the total Hilbert space has a dimension of 155,117,520. In this extremely large Hilbert space, we find 660 zero modes for $H^{(2,2)}$ (as we should). If we fix $w_1$,\ldots, $w_4$ and compute the zero modes of $H(w_1,\ldots,w_4)$$\downharpoonright$$_{{\cal H}_0}$, we find two zero modes (as we should). Fig.~\ref{example2} shows the density and the pair amplitudes corresponding to an orthogonal splitting of the zero modes space. The probes were fixed in a tetrahedral configuration for maximum separation. The orthogonal splitting of the zero modes space was generated by adding an infinitesimal, degeneracy lifting, Coulomb interaction to $H(w_1,\ldots,w_4)$.  The position of the probes is indicated by red dots. One can see that, at the probe locations, the electron density is finite while the pair amplitude is exactly zero. Also, judging by these plots, one could say that the quasi-holes are separated. We should also point out that  plots for the two zero modes look very similar, which is again the signature of the topological degeneracy.

\begin{figure}
  \includegraphics[width=8.6cm]{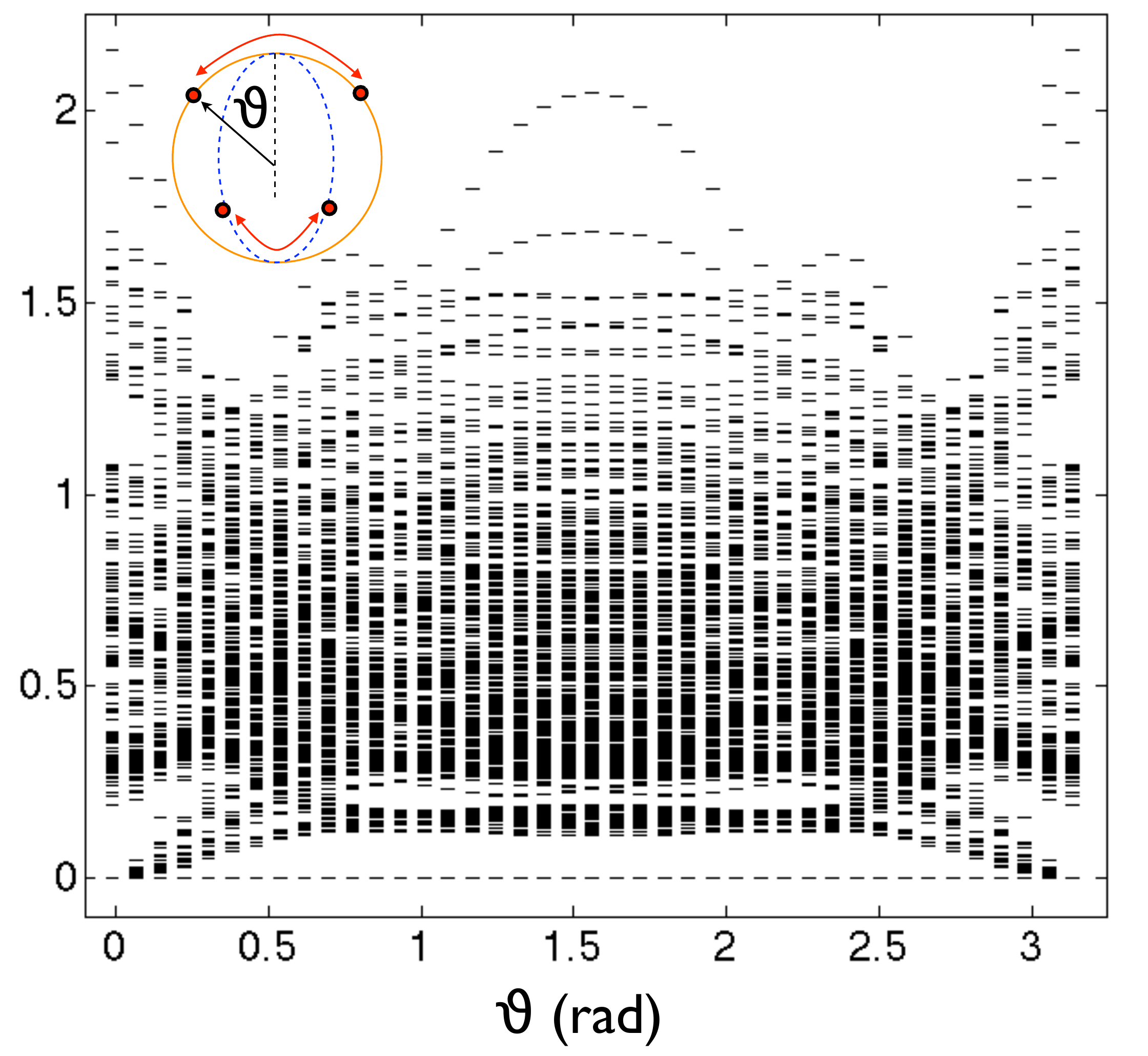}\\
  \caption{(Color online) The spectrum of $H(w_1,\ldots,w_4)$$\downharpoonright$$_{{\cal H}_0}$ as a function of the probes' position. The probes were moved along the for meridians: $(\theta,\phi=0)$, $(\theta,\phi=\pi)$, $(\pi-\theta,\phi=\pi/2)$, $(\pi-\theta,\phi=-\pi/2)$ (see inset). The results correspond to $N_{el}=14$ and $N_\phi = 27$.}
  \label{EigVdistance2}
\end{figure}

We continue our analysis of the case $N_\phi$=$N_\phi^0$+2 by plotting the spectrum of $H(w_1,\ldots,w_4)$$\downharpoonright$$_{{\cal H}_0}$ for different probe locations. For this, we moved the probes continuously along the paths shown in the inset of Fig.~\ref{EigVdistance2}, which can be described by: $w_1(\theta)$=($\theta$,$\phi$=0), $w_2(\theta)$=($\theta$,$\phi$=$\pi$), $w_3(\theta)$=($\pi$-$\theta$,$\phi$=$\pi$/2) and $w_4(\theta)$=($\pi$-$\theta$,$\phi$=-$\pi$/2). Fig.~\ref{EigVdistance2} plots the eigenvalues of $H(w_1,\ldots,w_4)$$\downharpoonright$$_{{\cal H}_0}$ as functions of $\theta$, for $N$=14 and $N_\phi$=27 (the graph becomes extremely busy if larger systems are used). For $\theta$ different from 0 or $\pi$, one can see in Fig.~\ref{EigVdistance2} the doubly degenerate zero energy level, separated by an energy gap from the rest of the spectrum (when compared with other energy separations visible in the graph). The gap becomes independent of the positions of the probes when the probe separation becomes larger than approximately 4 magnetic lengths. As the probes come closer to each other, the gap decreases and, at the intersection points $\theta$=0 and $\pi$, the gap is completely closed, leading to an abrupt increase in the level degeneracy.

\section{Braiding the Moore-Read states}

\subsection{General considerations.} 

Let us consider the general situation when the number of flux has been increased by $n$ units, $N_\phi$=$N_\phi^0$+$n$. We use the pinning Hamiltonian $H(w_1,\ldots,w_{2n})$$\downharpoonright$$_{{\cal H}_0}$ given in Eq.~(\ref{pinning}) to control the position of the anyons. Given that each $w_i$ is a 2-dimensional variable, the Hamiltonian depends, parametrically, on $4n$ coordinates. We use the symbol ${\bf x}$ to denote these coordinates and the shorthand $H({\bf x})$ for $H(w_1,\ldots,w_{2n})$$\downharpoonright$$_{{\cal H}_0}$. 

Let us consider an arbitrary closed path in the parameter space: $\ell  \rightarrow  {\bf x}(\ell)$ (${\bf x}_0$ =${\bf x}(0)$), where we use the length $\ell$ to parametrize the loop. Now imagine that we move along this path with constant velocity $v$. One would like to study the time evolution $U(t)$ generated by the time dependent Hamiltonian $H(t) \equiv H({\bf x}(vt))$:
\begin{equation}
	i\partial_t U(t) = H(t) U(t), \  U(0)=I,
\end{equation}
in the adiabatic limit, i.e. in the limit when the motion along the path is infinitely slow.

\begin{figure}
  \includegraphics[width=6.0cm]{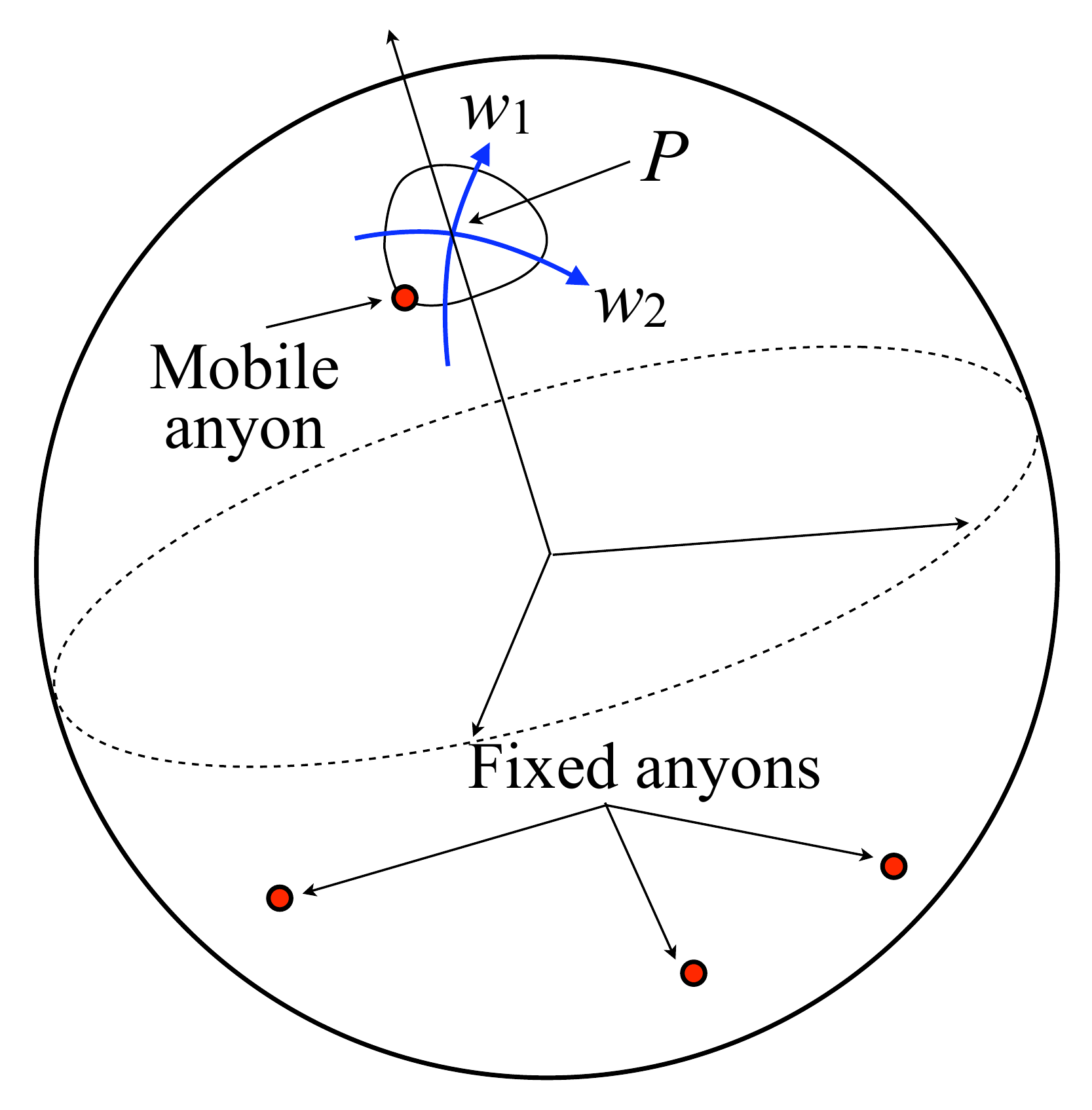}
  \caption{(Color online) The figure illustrates a configuration in which all anyons are kept fixed except one, which is moved on the sphere. To compute different quantities at an arbitrary point $P$ of the sphere, we introduce a local coordinate system $(w_1,w_2)$ as described in the text.}
  \label{braiding}
\end{figure}

If we denote by $P_{\bf x}$ the projector onto the zero modes space of $H({\bf x})$, then the classic Adiabatic Theorem says that:\cite{Nenciu:1981kx,Avron:1987fk,Nenciu:1993uq}
\begin{equation}\label{adapprox}
U(t)P_{{\bf x}_0}=U_a(t)P_{{\bf x}_0}+o(v,t),
\end{equation}
where $U_a(t)$ is the adiabatic propagator, i.e. the unique solution of the following system of equations:\cite{Avron:1987fk}
\begin{equation}\label{AdEvol}
\begin{array}{c}
i\partial_t U_a(t) = (H(t)+i[\partial_t P_{\bf x},P_{\bf x}]) U_a(t), \medskip \\ 
U_a(0)=I.
\end{array}
\end{equation}
As long as the energy gap between the zero energy of the anyons and the energies of  the excited states remains open, the adiabatic evolution is a good approximation of the true evolution, when restricted to the zero modes space. In fact, the two become identical if the velocity $v$ is infinitely small. We want to point out that there are rigorous estimates (see for example Refs.~\onlinecite{Nenciu:1981kx,Avron:1987fk,Nenciu:1993uq}) of the errors $o(v,t)$ that are made when one approximates $U(t)$ by $U_a(t)$, i.e. of the non-adibatic effects. Since the braids corresponding to the elementary TQC logic gates are quite long,\cite{Bonderson:2006qf} these estimates may play an important role in the field.   

The adiabatic evolution satisfies the following fundamental relation:
\begin{equation}\label{AdInv}
P_{\bf x} = U_a({\bf x})P_{{\bf x}_0}U_a({\bf x})^{-1}.
\end{equation}
This equation tells us that the zero modes space is mapped into itself by the adiabatic evolution. This mapping can be computed by solving Eq.~(\ref{AdEvol}), which tells us how the zero modes space is transported during the adiabatic evolution. One can put all these into a geometric perspective, by defining the following 1-form (the adiabatic connection):
\begin{equation}\label{AdConn}
\dd a=\frac{1}{v}H({\bf x}) d\ell+i[\dd P_{\bf x},P_{\bf x}],
\end{equation}
in which case $U_a$ can be viewed as the solution of the following differential equation:
\begin{equation}\label{adiabaticg}
i\dd U_a=\dd a \:  U_a, \ \ U_a({\bf x}_0)=I.
\end{equation}
Thus, by fixing the speed $v$, we have eliminated the time. The quantum algorithms will use the adiabatic unitary transformations $U_a(\Gamma)$, $U_a(\Gamma')$, \ldots, resulted from taking the probes along certain closed paths $\Gamma$, $\Gamma'$, \ldots.

Assume that, for each configuration of the anyons (equal to say that for each ${\bf x}$), we chose a basis set (a gauge) $\psi_1({\bf x}),...,\psi_{D}({\bf x})$ in the $D$=$2^{n-1}$ dimensional zero mode space. We use $\vec{\psi}({\bf x})$ to denote the vector of components $\psi_1({\bf x}),...,\psi_D({\bf x})$. Due to the fundamental property of the adiabatic evolution Eq.~(\ref{AdEvol}), there exists a unitary $D$$\times$$D$ matrix $\hat{W}({\bf x})$ such that
\begin{equation}
	U_a({\bf x}) \vec{\psi}({\bf x}_0)=\hat{W}({\bf x})\vec{\psi}({\bf x}).
\end{equation}
The Wilczeck-Zee connection is given by:\cite{Wilczek:1984bs}
\begin{equation}
\dd \hat{A} = i\hat{W}({\bf x})^{-1}\dd \hat{W}({\bf x}),
\end{equation}
which takes the classical form:
\begin{equation}
\dd A_{ij}({\bf x}) = -\frac{1}{v}H({\bf x})_{ji}d\ell-i\langle \psi_j({\bf x}),\dd \psi_i({\bf x}) \rangle,
\end{equation}
where $H({\bf x})_{ji}$ are the matrix elements of the Hamiltonian in the chosen basis set. Since we are dealing with zero modes, the first term in the right hand side, above, is identically zero.

The $D$$\times$$ D$ unitary matrix $\hat{W}({\bf x})$ implements the adiabatic evolution in the invariant subspaces $P_{\bf x}{\cal H}_0$ and it can be computed as the unique solution of the differential equation:
\begin{equation}\label{mon}
i\dd \hat{W}({\bf x}) = \hat{W}({\bf x}) \ \dd \hat{A}, W({\bf x}_0)=I
\end{equation}
which is to be integrated along the braiding path. For a loop $\Gamma$ that starts and ends at ${\bf x}_0$, we solved this equation numerically, by considering a large number of points along the loop, ${\bf x}_0$,...,${\bf x}_K$, and constructing the monodromy:
\begin{equation}
\hat{W}_\Gamma=P_{{\bf x}_0}P_{{\bf x}_1}P_{{\bf x}_{2}}...P_{{\bf x}_K}P_{{\bf x}_0}.
\end{equation}
This amounts to finding the null space of the pinning Hamiltonian for each ${\bf x}_k$. If we define $W_\Gamma(k) = P_{{\bf x}_0}P_{{\bf x}_{2}}...P_{{\bf x}_k}$, then
\begin{eqnarray}
 \hat{W}_\Gamma (k+1)-W_\Gamma(k) =\hat{W}_\Gamma(k) (P_{{\bf x}_{k+1}}-P_{{\bf x}_k})P_{{\bf x}_k} \nonumber \\
- \hat{W}_\Gamma(k) P_{{\bf x}_{k+1}}(P_{{\bf x}_{k+1}}-P_{{\bf x}_k}),
\end{eqnarray}
which is the finite difference version of
\begin{equation}
i\dd \hat{W}_\Gamma(x)=i\hat{W}_\Gamma(x)[\dd P_{\bf x}, P_{\bf x}],
\end{equation}
which is the same as Eq.~(\ref{mon}). The numerically calculated $\hat{W}_\Gamma$ matrix becomes a unitary matrix only in the limit when the number of discrete points goes to infinity. To quantify how much does $\hat{W}_\Gamma$ deviate from a unitary matrix, we compute the absolute value of the determinant of $\hat{W}_\Gamma$, which is compared to the value of 1, appropriate for a unitary matrix. In all the calculations presented in this paper, $K$ was chosen large enough so that $|\det \hat{W}_\Gamma|$=0.999 or better. This is a measure of how well converged are our numerical calculations.

\subsection{The Quantum Geometry of the zero modes states}

We can endow the zero modes with a curvature. The curvature form associated with the adiabatic connection is given by:
\begin{equation}
d\hat{F}=\left \{ \partial_\mu \hat{A}_\nu - \partial_\nu \hat{A}_\mu - i [\hat{A}_\mu , \hat{A}_\nu ] \right \} dx^\mu \wedge dx^\nu.
\end{equation}
The explicit expressions of its coefficients are:
\begin{equation}
(\hat{F}_{\mu \nu})_{ij}({\bf x})= 2\text{Im}\langle \partial_\mu \psi_j ({\bf x}), [1-P_{\bf x}]\partial_\nu \psi_i ({\bf x}) \rangle. 
 \label{curv}
\end{equation}

Besides the adiabatic connection and curvature, we can endow the parameter space with an intrinsic metric tensor, which we will refer to as the quantum metric tensor. First of all, we can introduce the following quantum distance:\cite{Marzari:1997ys}
\begin{equation}
d^q({\bf x},{\bf x}') = \| P_{\bf x} - P_{{\bf x}'} \| _{HS},
\end{equation}
where HS means the Hilbert-Schmidt norm. This distance is at least second order differentiable in the coordinates ${\bf x}$ and ${\bf x}'$ and for this reason we can generate the quantum metric tensor $g_{\mu \nu}$ via a Taylor expansion:
\begin{equation}
d^q({\bf x},{\bf x}+\delta{\bf x}) = \frac{1}{2}g^q_{\mu\nu}({\bf x}) \delta x^\mu \delta x^\nu +o(\delta {\bf x}^3).
\end{equation}
The coefficients of the quantum metric tensor are given by the classical expression:\cite{Marzari:1997ys}
\begin{equation}\label{qmetric}
g^q_{\mu \nu}({\bf x}) =2 \text{Re} \sum_i \langle \partial_\mu \psi_i({\bf x}) ,[1-P_{\bf x})]\partial_\nu \psi_i({\bf x}) \rangle.
\end{equation}

\begin{figure}
  \includegraphics[width=8.6cm]{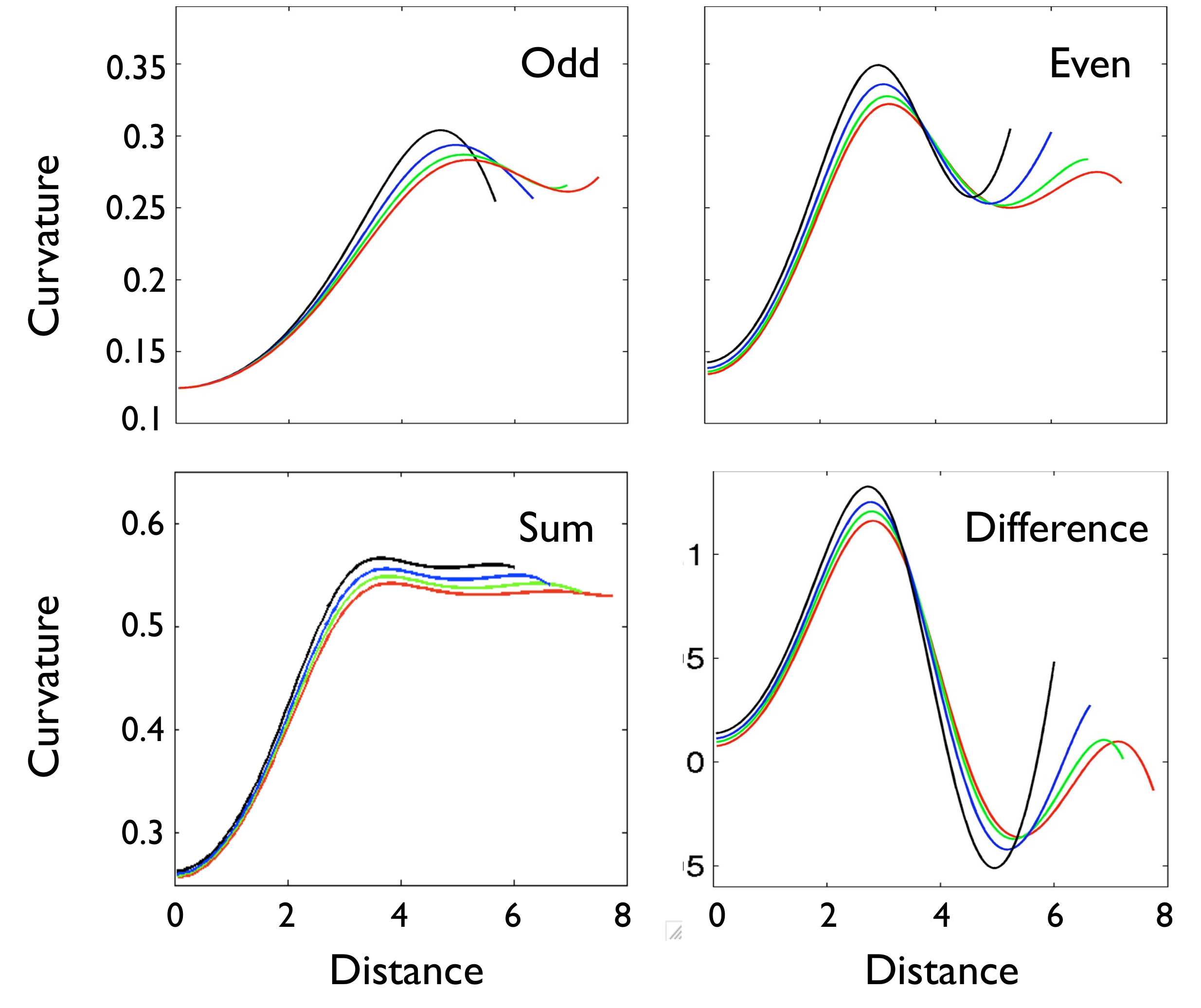}\\
  \caption{(Color online.) The adiabatic curvature $F(P)$ obtained by moving one anyon while keeping the other fixed. Left-upper panel shows the results for odd number of electrons: $N/N_\phi$=9/16, 11/20, 13/24, 15/28 and the right-upper panel shows the results for even number of electrons: $N/N_\phi$=10/18, 12/22, 14/26, 16/30. The lower-left and lower-right panels show the sum and the difference between the odd and even results, respectively. For example, we added and subtracted the result for $N/N_\phi$=10/18 and $N/N_\phi$=9/16, and then the results for $N/N_\phi$=12/22 and $N/N_\phi$=11/20, etc..}
  \label{Curvature1Flux}
\end{figure}

In the following, we demonstrate an interesting relation between the adiabatic curvature and the quantum metric tensor. We fix all anyons, except one, in which case the parameter space becomes 2-dimensional.  A point in this parameter space describes the position of the itinerant anyon on the sphere.  To compute the coefficient of the curvature form at an arbitrarily chosen point $P$ on the sphere, we introduce a local coordinate system by using the complex coordinate  $w=2R \tan \frac{\theta}{2} e^{-i\phi}$, where $(\theta,\phi)$ are the usual spherical parameters when the North pole is  fixed at $P$. We take $w_1$=$\text{Re}\{w\}$ and $w_2$ =$ \text{Im}\{w\}$ as the two independent variables (see Fig.~\ref{braiding}). In these local coordinate system, the curvature form becomes
\begin{equation}\label{curvatureForm}
dF = \hat{F}(w)dw^1 \wedge dw^2,
\end{equation}
where $\hat{F}$ is a $D$$\times$$D$ matrix. At $P$, in the same system of coordinates, we can prove the following general fact:
\begin{equation}\label{fact}
g^q_{\mu \nu}(P) = \text{Tr} \{ \hat{F}(P)\} \ g^0(P)_{\mu \nu},
\end{equation}
where $g^0$ is the standard metric of the sphere. This shows that the quantum metric and the standard metric of the sphere are related by a conformal transformation, which involves the trace of the curvature tensor.

\begin{figure}
  \includegraphics[width=8.6cm]{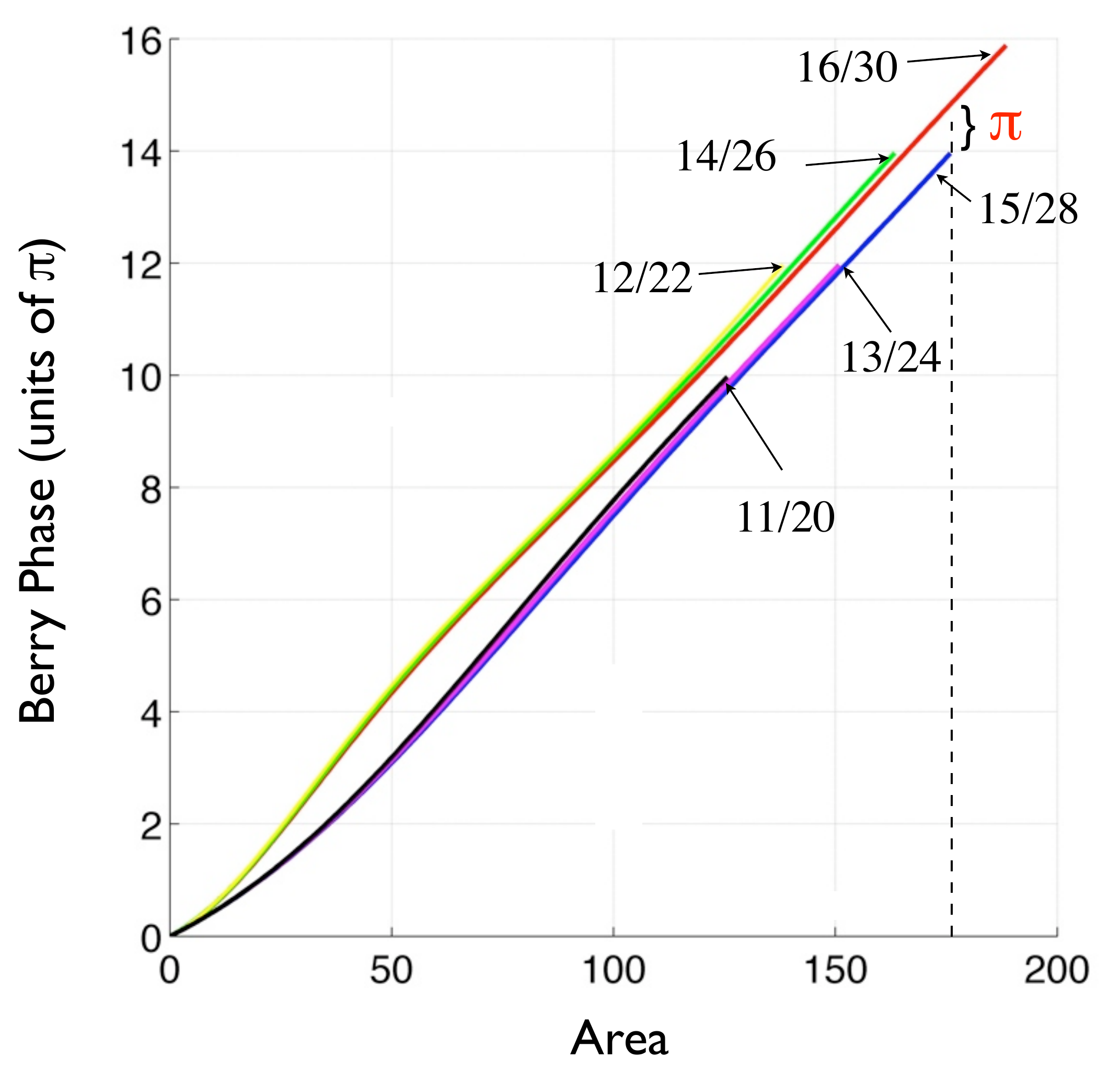}\\
  \caption{(Color online.) The Berry phase accumulated by an anyon when moved along a path $\theta$=const, with the other anyon fixed at the North pole. Different lines refer to different system sizes that are marked in the diagram with their corresponding $N/N_\phi$ numbers. A point on a particular line represents one braiding path, which is labeled by the area enclosed by the path, shown on the horizontal axis. }
  \label{Phase1Flux}
\end{figure}

\begin{figure*}
  \includegraphics[width=17.2cm]{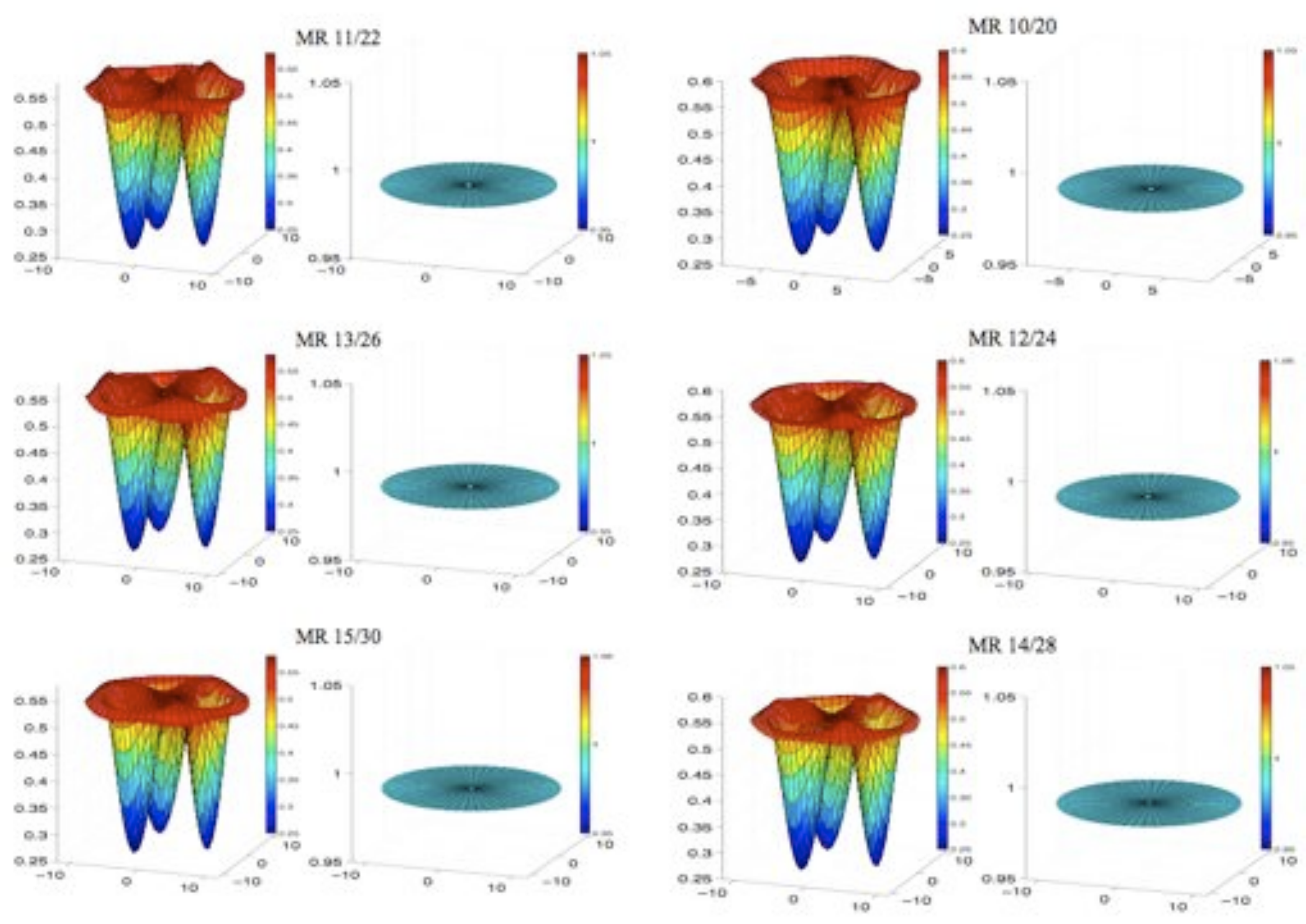}\\
  \caption{(Color online) Plots of Tr$\hat{F}(P)$ for different system sizes (each panel is marked with the corresponding $N/N_\phi$). For each size, we show Tr$\hat{F}(P)$ calculated with the standard metric tensor (left) and with the quantum metric tensor (right).}
  \label{AbelianCurv}
\end{figure*}

We start the proof of Eq.~(\ref{fact}). The $2^{n-1}$ many body wavefunctions $\Psi_i$ of the zero modes depend parametrically on the position $w$ of the mobile anyon and take the following general form:\cite{Read:1996nx}
\begin{equation}
\Psi_i(w) = {\cal A}_i(w) \tilde{\Psi}_i(w), \ i=1,\ldots,2^{n-1}
\end{equation}
where the crucial fact is that the vectors $\tilde{\Psi}_i(w)$ are all analytic of $w$.  ${\cal A}_i(w)$ are normalization factors that are not analytic of $w$. We have:
\begin{equation}
\text{Tr}\{ \hat{F}(w)\} = 2 \text{Im} \sum\limits_i \langle \partial_1 \Psi_i(w),[1-P_w]\partial_2 \Psi_i(w) \rangle.
\end{equation} 
Due to the presence of $1-P_w$, we can ignore the action of the derivatives on the normalization constants ${\cal A}_i(w)$, which then can be pulled out of the scalar product as below:
\begin{equation}
\begin{array}{c}
\text{Tr} \hat{F}(w) = 2  \sum\limits_i |{\cal A}_i(w)|^2 \times \medskip \\
 \text{Im}\langle \partial_1 \tilde{\Psi}_i(w),[1-P_w]\partial_2 \tilde{\Psi}_i(w) \rangle.
\end{array}
\end{equation}
Next we use the fact that $\tilde{\Psi}_i$ are analytic functions of $w$, i.e. they are functions of the single variable $w$ rather than of two independent variables $w_1$ and $w_2$, in which case:
\begin{equation}
\begin{array}{c}
\text{Tr} \hat{F}(w) = 2  \sum\limits_i |{\cal A}_i(w)|^2 \times \medskip \\
\text{Im}\{ \langle \partial_w \tilde{\Psi}_i(w),[1-P_w]\partial_w \tilde{\Psi}_i(w) \rangle(\partial_1w)^*\partial_2 w\}.
\end{array}
\end{equation}
The scalar product is real, which allows us to easily compute the imaginary part, resulting in:
\begin{equation}
\begin{array}{c}
\text{Tr} \hat{F}(w) = 2  \sum\limits_i |{\cal A}_i(w)|^2 \times \medskip \\
  \langle \partial_w \tilde{\Psi}_i(w),[1-P_w]\partial_w \tilde{\Psi}_i(w) \rangle.
\end{array}
\end{equation}
If one repeats exactly the same arguments for the quantum metric tensor given in Eq.~\ref{qmetric}, the conclusion will be that:
\begin{equation}
\begin{array}{c}
g^q_{\mu \nu}(w) = 2  \sum\limits_i |{\cal A}_i(w)|^2 \medskip \\
\times  \langle \partial_w \tilde{\Psi}_i(w),[1-P_w]\partial_w \tilde{\Psi}_i(w) \rangle \delta_{\mu \nu}.
\end{array}
\end{equation}
Since $w_1$ and $w_2$ coincide with the geodesic coordinates of the sphere at $P$, i.e. the coordinates in which the metric tensor becomes the identity matrix when evaluated at $P$: $g^0(P)_{\mu \nu}=\delta_{\mu \nu}$, the proof of Eq.~\ref{fact} is completed.

\section{Quantum geometry for one flux added.} 

In this case we have 2 anyons and the zero modes space is 1-dimensional. We keep one anyon fixed and move the other along different braiding paths. The monodromy of any path $\Gamma$ can be computed by integrating the curvature:
\begin{equation}\label{curv}
\hat{W}_\Gamma = e^{i\int_{S_\Gamma} dF},
\end{equation}
where $S_\Gamma$ is the surface enclosed by $\Gamma$. Thus, if we map the curvature, we can easily compute the monodromy of any arbitrary path.

We compute the coefficient $F$ of the curvature form (see Eq.~\ref{curvatureForm}) at a point $P$ of the sphere from:
\begin{equation}\label{curv}
F(P)=\lim_{S_\Gamma\rightarrow 0}\frac{\hat{W}_\Gamma-1}{iS_\Gamma},
\end{equation}
where $\Gamma$ is a small path around $P$. The monodromy is computed as explained in the previous section and we obtain the limit by considering paths of decreasing radius. The value of $F$ computed this way coincides with the coefficient of the curvature in the local coordinate system $(w_1,w_2)$ introduced in the previous section. The monodromy Eq.~\ref{curv}, however, is independent of the coordinates used to compute the curvature form.

The upper panels in Fig.~\ref{Curvature1Flux} plot $F(P)$ as a function of the distance from $P$ to the position of the fixed anyon. The two upper panels refer to odd and even numbers of electrons. The lower two panels show the sum and the difference between the results for odd and even number of electrons. Since one is interested in the thermodynamic limit, we plot sequences of curves for an increasing number of electrons. By comparing the curves for these sequences, one can determine how fast is the thermodynamic limit achieved. 

\begin{figure*}
  \includegraphics[width=15cm]{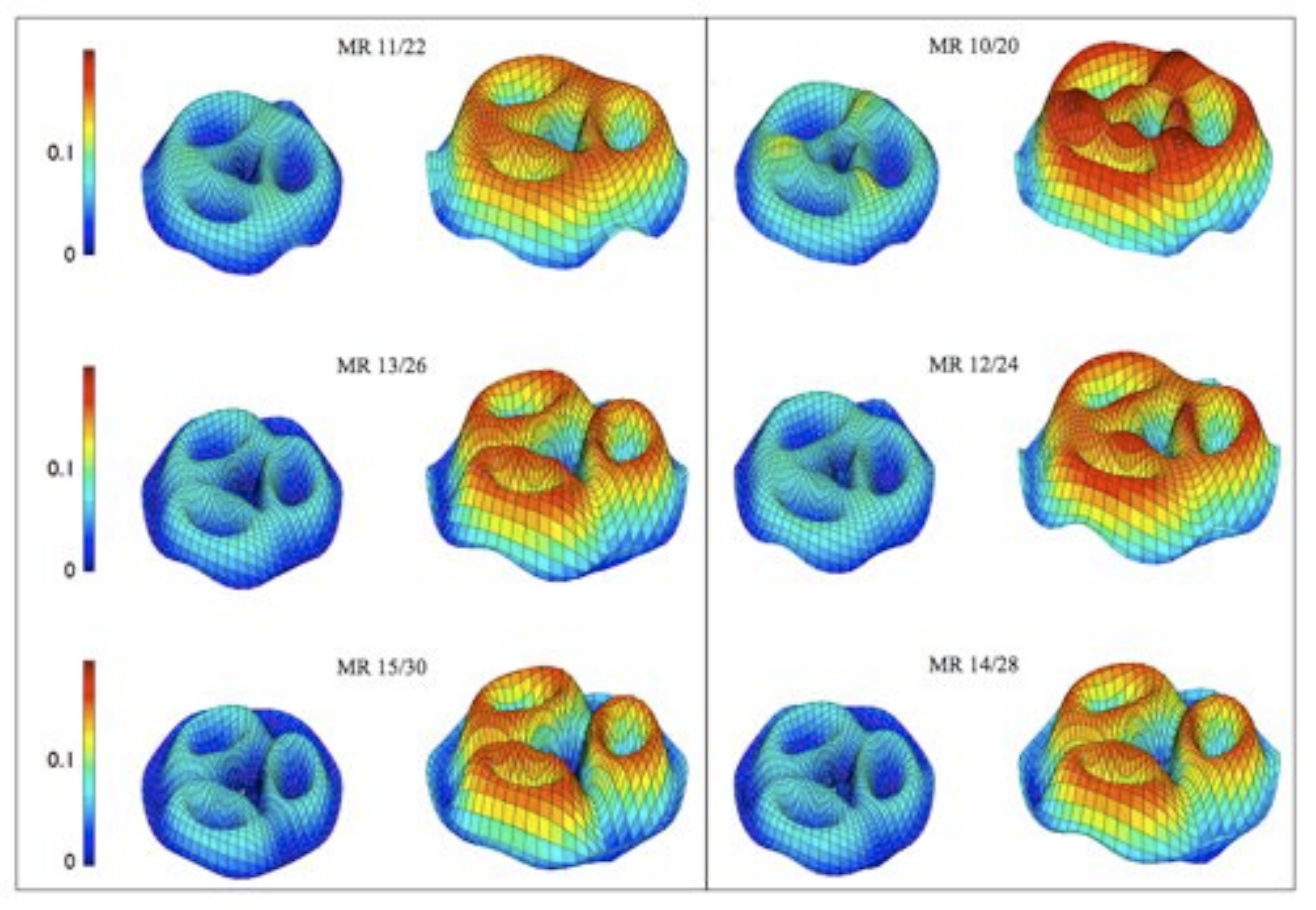}\\
  \caption{(Color online) Plots of the amplitude of the non-Abelina part of the curvature $|{\bf f}(P)|$ for different system sizes (each panel is marked with the corresponding $N/N_\phi$). For each size, we show $|{\bf f}(P)|$ calculated with the standard metric tensor (left) and with the quantum metric tensor (right).}
  \label{NonAbelianCurv}
\end{figure*}

As one can clearly see, the curves in the top panels of Fig.~\ref{Curvature1Flux} go asymptotically with the distance towards a constant value, which is precisely equal to the quasi-hole charge $e^*$=$e$/4 (when working on the sphere, there is a small correction to this value, correction that goes to zero as the size of the sphere is increased). The most remarkable thing about Fig.~\ref{Curvature1Flux} is that the curvatures for odd/even number of electrons have different thermodynamic limits, as one can clearly see by inspecting the difference between the two, plotted in the lower-right panel.

From the adiabatic curvature we compute the Berry phases accumulated (i.e. the exponent in Eq.~\ref{curv}) as we move one anyon along different paths  of the form $\theta$=ct., while keeping the other anyon fixed at the North Pole of the sphere. Fig.~\ref{Phase1Flux} shows the Berry phase as function of the area enclosed by the path. We use this figure to draw several important conclusions. First, we point out that the Berry phase plotted in this figure includes also the Aharonov-Bohm phase due to the magnetic flux $\phi_B$: $\psi_{AB}$=$e^*\phi_B$. Thus it is expected that the total Berry phase, as a function of the enclosed area, to go asymptotically to a linear curve, a feature that is obviously present in Fig.~\ref{Phase1Flux}. The slope of the asymptotic part of the curve is equal to the charge $e^*$=$e/4$ of the anyons. 

Excepting the length of the lines, the graphs for different system sizes look very similar, implying that the thermodynamic limit is achieved very fast. We expect the graphs for the larger systems to represent the thermodynamic limit with a high degree of accuracy. Looking at the largest systems, we can see that, for large enclosed areas, the curves representing the Berry phase for systems with even number of electrons are shifted upward by $\pi$ relative to the curves representing the Berry phase for systems with odd number of electrons. This is a topological effect, since the shift is independent of the area enclosed by the braiding loop or of the shape of the loop, as long as the loop is large enough. {\it The finding is in total agreement with the SO spinorial representation of the braid group derived in Ref.~\onlinecite{Nayak:1996mb} on the basis of underling CFT structure of the Moore-Read state.} {\it This effect was also considered the most direct signature of the Non-Abelian statistics for the Moore-Read state}.\cite{Ivanov:2001gd,Tserkovnyak:2003vn} 

We conclude this subsection with the observation that the coefficient $F$ depends on how we compute the area enclosed by $\Gamma$ in Eq.~\ref{curv}, more precisely on what metric tensor is used when computing the area enclosed by the loops. The plots shown in Fig.~\ref{Curvature1Flux} were obtained with the standard metric of the sphere. We have repeated the calculations using the quantum metric tensor Eq.~(\ref{qmetric}) instead, in which case we have to re-scale the coefficient:
\begin{equation}
F^q(w) = F(w)/\sqrt{\det g^q_{\mu \nu}(w)}.
\end{equation}
According to the our previous analytic prediction (see Eq.~\ref{fact}), $F^q(P)$ should be identically 1. We numerically checked this prediction in the following way.
The quantity that is readily available in the numerical calcualtions is the quantum distance. We can compute the determinant of the quantum metric tensor by considering a circle of radius $\rho$ (in the standard metric of the sphere) centered at $P$ and calculate the quantum distance between $P$ and the points of this circle. If $d^q_M$ and $d^q_m$ denote the maximum, respectively minimum quantum distance to the points of the circle, then
\begin{equation}
\det g^q_{\mu \nu}(P)=\lim_{\rho \rightarrow 0} \frac{(d_m^qd_M^q)^2}{\rho^4}.
\end{equation}
The computation of the determinant is done simultaneously with the calculation of the curvature, which also requires the walk on the same circle. The numerical calculations give a direct confirmation that $F^q(P)$=1 for all points of the sphere.

\section{Quantum geometry for two fluxes added} 

In this case we have 4 anyons and the zero modes space is 2-dimensional. We will keep 3 anyons fixed and move the forth one along different braiding paths.

For the Non-Abelian case, there is no simple Stokes theorem,\cite{Diakonov:2001zr,Faber:2000ly} which means the monodromy can not be simply computed from the curvature as we did for the previous case. Even so, mapping the curvature provides a clear picture of the non-comutative and topological properties of the states. 

The parameter space remains 2-dimensional. As before, a point in this parameter space indicates the position of the mobile anyon.  Thus $dF$=$\hat{F} dw^1$$\wedge$$dw^2$, but $\hat{F}$ is now a 2$\times$2 matrix. We compute $\hat{F}(P)$ using the same algorithm (see Eq.~\ref{curv}). Using the Pauli's matrices, $\sigma_i$, i=1, 2, 3, $\hat{F}(P)$ can be uniquely decomposed as:
\begin{equation}\label{sigmarep}
\hat{F}(P) = f_0(P) + {\bf f}(P)\cdot\text{\boldmath{$\sigma$}},
\end{equation}
where $f_0(P)$=$\frac{1}{2}$Tr$\hat{F}(P)$ and ${\bf f}(P)$ is a 3-component vector. We will refer to $f_0$ as the Abelian and to ${\bf f}$$\cdot$$\text{\boldmath{$\sigma$}}$ as the non-Abelian part of the curvature. It is important to notice how different quantities behave when changing the gauge, i.e. the basis in the 2-dimensional zero modes space. We have: $f_0(P)$  is gauge independent; the magnitude of ${\bf f}(P)$ is gauge independent; the orientation of ${\bf f}(P)$ is gauge dependent. 

Fig.~\ref{AbelianCurv} shows plots of Tr$\hat{F}(P)$ for different system sizes. For each size, we show Tr$\hat{F}(P)$ calculated with the standard and with the quantum metric tensor. The numerics confirm again the theoretical prediction that Tr$\hat{F}(P)$=1.

\begin{figure}
  \includegraphics[width=7.6cm]{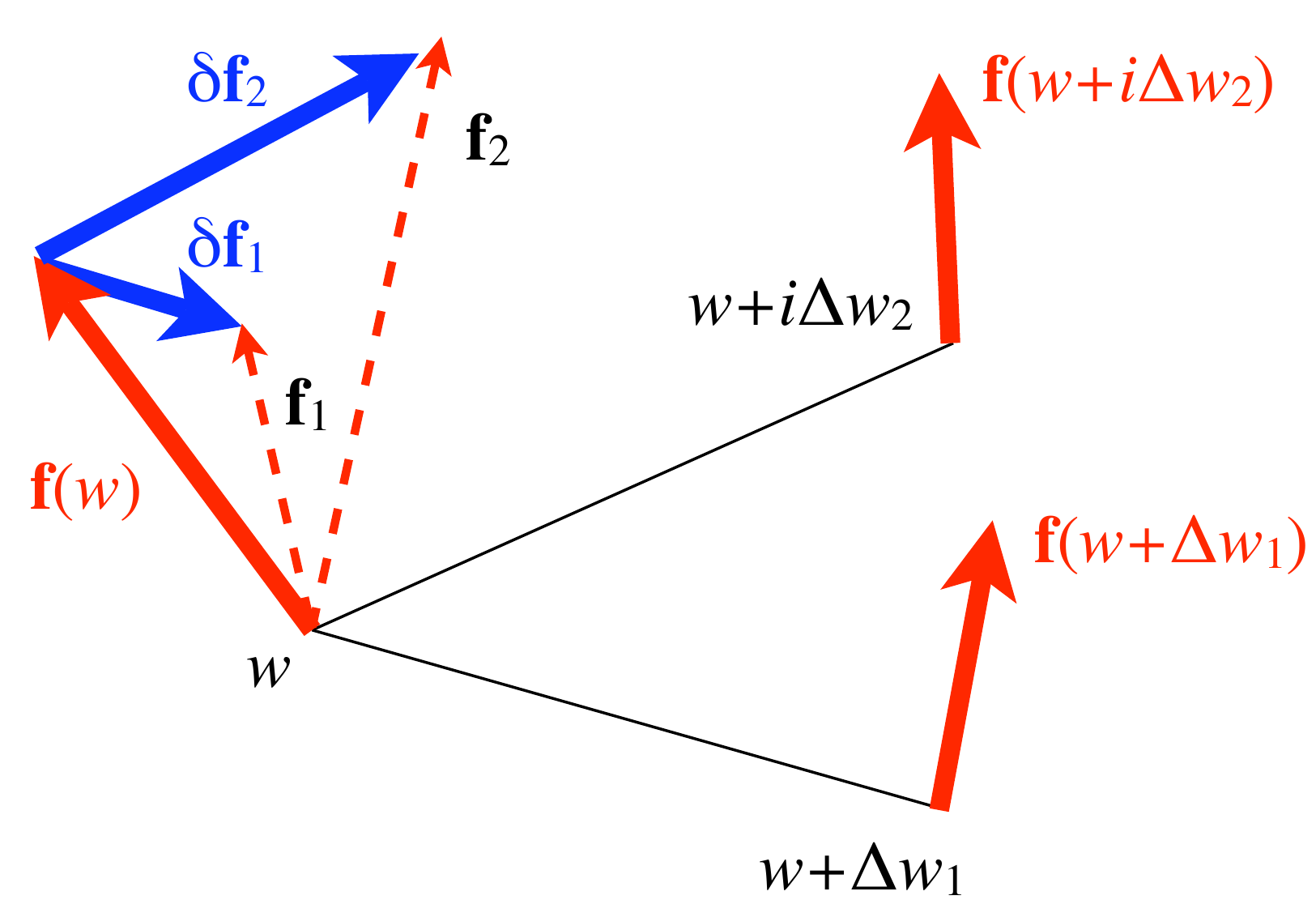}\\
  \caption{(Color online) A diagram of the parallel transport used in the calculation of the Twist density $\rho_{\text{\tiny{TW}}}$.}
  \label{ptransp}
\end{figure}

To demonstrate that the braid group is non-comutative, we need first to show that ${\bf f}(P)$ is non-zero. This, however, is not enough. We need also to rule out the existence of a particular gauge in which the adiabatic connection becomes diagonal at every point of the parameter space. If such a gauge exists, the fiber bundle of the zero-modes degenerates into a trivial $U(1)$$\times$$U(1)$ fiber bundle, in which case there will be two 1-dimensional fibers that do not mix during the adiabatic braiding. Consequently, all monodromies will take a diagonal form in this gauge and they will commute with each other.

As already mentioned, the magnitude of ${\bf f}(P)$ is gauge independent. Thus we can see if ${\bf f}(P)$ is zero or not by simply plotting its magnitude, which is shown in Fig.~\ref{NonAbelianCurv} for different sizes. The graph clearly demonstrates that ${\bf f}(P)$ is non-zero and it appears to be concentrated near the positions of the fixed anyons. We will further discuss Fig.~\ref{NonAbelianCurv} in the next Section.

Next, we introduce a scalar function which we call the Twist density $\rho_{\text{\tiny{TW}}}$, which gives a measure of how much are the fibers twisted during the adiabatic parallel transport. We start the construction from the following 2-form:
\begin{equation}
d\hat{\rho} =[D_\mu \hat{F},D_\nu \hat{F}] \ dw_\mu \wedge dw_\nu,
\end{equation}
where $D_\mu$ denotes the covariant derivative corresponding to the adiabatic connection and $[,]$ denotes the usual commutator. This form is invariant to coordinate transformations, thus well defined. The coefficients of this form are $2\times 2$ matrices. In the special case of a two dimensional parameter space, the form reduces to:
\begin{equation}
d\hat{\rho} =[D_1 \hat{F},D_2 \hat{F}] \ dw_1 \wedge dw_2,
\end{equation}
Based on the above observations, we construct the following 2-form,
\begin{equation}
d\rho_{\text{\tiny{TW}}}  \equiv \sqrt{\det[D_1 \hat{F},D_2 \hat{F}]} \ dw_1 \wedge dw_1,
\end{equation}
whose coefficient is a pseudo-scalar function. Since $d\hat{\rho}$ is invariant to coordinate transformations, $d\rho_{\text{\tiny{TW}}}$ is also invariant and hence well defined. The density
\begin{equation}\label{def1}
\rho_{\text{\tiny{TW}}}=\sqrt{\det [D_1 \hat{F},D_2 \hat{F}]}
\end{equation}
is gauge invariant and it is identically zero for trivial fiber bundles, in particular for $U(1)$$\times$$U(1)$ fiber bundle over our parameter space. Thus, if we show that $\rho_{\text{\tiny{TW}}}$ is non-zero, that will be equivalent to demonstrating that the zero modes fiber bundle is non-trivial.

\begin{figure*}
  \includegraphics[width=15cm]{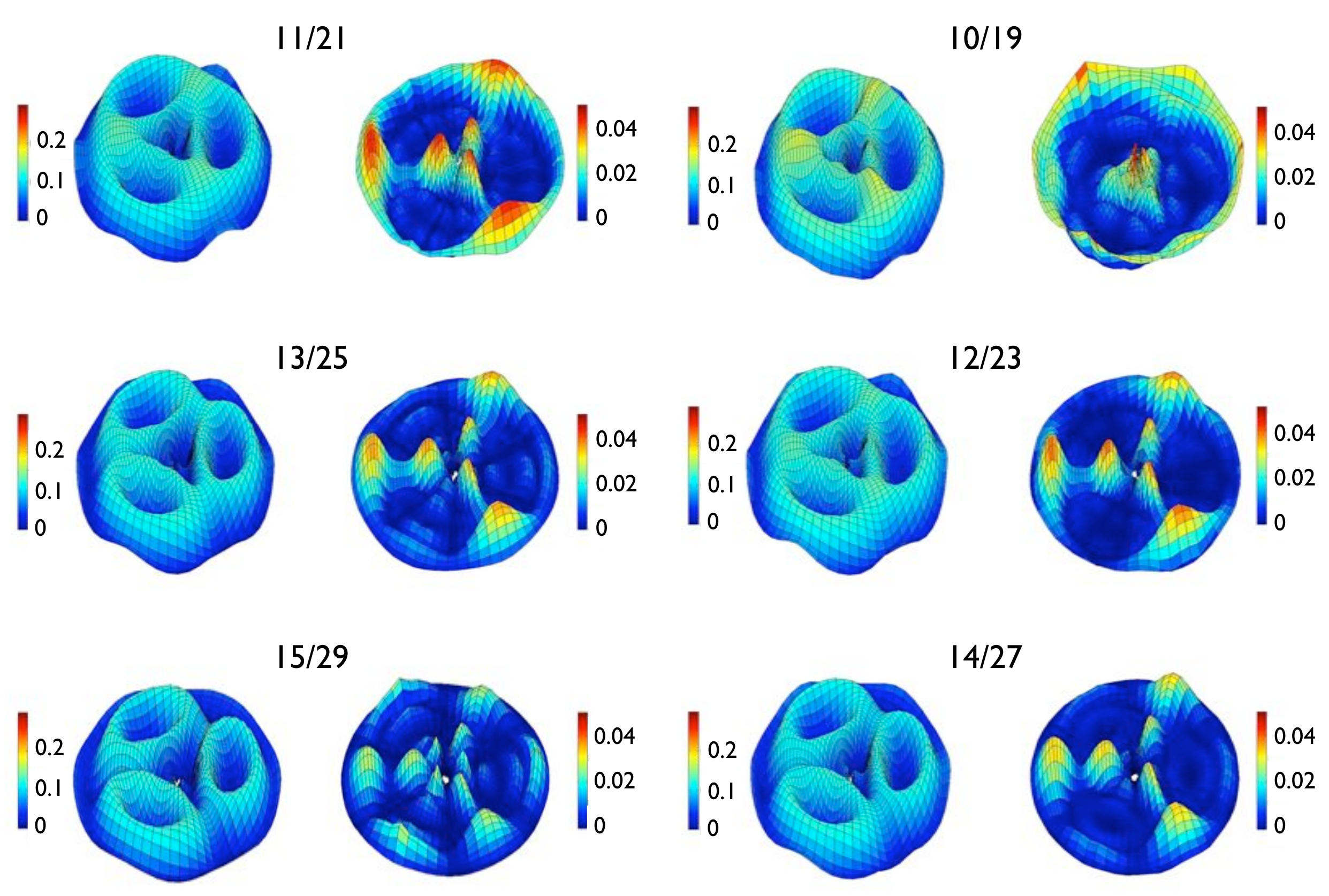}\\
  \caption{(Color online) Plots of $\rho_{\text{\tiny{TW}}}(P)$ for different system sizes (each panel is marked with the corresponding $N/N_\phi$). To see how $\rho_{\text{\tiny{TW}}}(P)$ relates to the adiabatic curvature, for each size, we show $|{\bf f}(P)|$ on the left and $\rho_{\text{\tiny{TW}}}(P)$ on the right.}
  \label{pontrjagin}
\end{figure*}

Let us now give the physical interpretation of our construction. For this we consider three points on the sphere: $w$, $w$+$\Delta w_1$ and $w$+$i\Delta w_2$, as in Fig.~\ref{ptransp}. Assume that we computed the curvature $\hat{F}$ at each of these points, in a pre-chosen, arbitrary gauge. The corresponding ${\bf f}$ (see Eq.~\ref{sigmarep}) vectors are gauge dependent. To compare the three vectors, we need to adiabatically transport them to the same point $w$. For this we compute the monodromies $\hat{W}_1$ for the path $w$+$\Delta w_1$$\rightarrow$$w$ and $\hat{W}_2$ for the path $w$+i$\Delta w_2$$\rightarrow$$w$ and then compute $\hat{F}_1$=$\hat{W}_1 \hat{F}(w$+$\Delta w_1) \hat{W}_1^{-1}$ and $\hat{F}_2$=$\hat{W}_2 \hat{F}(w$+i$\Delta w_2) \hat{W}_2^{-1}$. By decomposing $\hat{F}_{1,2}$ as in Eq.~\ref{sigmarep}, we obtain the parallel transport of ${\bf f}(w$+$\Delta w_1)$ and ${\bf f}(w$+$i\Delta w_2)$ to $w$. We denote them by ${\bf f}_{1}$ and ${\bf f}_{2}$, respectively. We now can ask if they are parallel. To quantify the answer to this question, we form the differences $\delta_1 {\bf f}$=${\bf f}_1$-${\bf f}$ and $\delta_2 {\bf f}$=${\bf f}_2$-${\bf f}$ and define
\begin{equation}\label{def2}
\rho_{\text{\tiny{TW}}} \equiv \lim_{\Delta w \rightarrow 0} \frac{| \delta_1 {\bf f} \times \delta_2{\bf f}|}{\Delta w_1 \Delta w_2}.
\end{equation}
This quantity is gauge invariant and measures how much are $\delta_1 {\bf f}$ and $\delta_2 {\bf f}$ deviating from being parallel. The two definitions of $\rho_{\text{\tiny{TW}}}$ given in Eqs.~(\ref{def1}) and (\ref{def2}) coincide. Fig.~\ref{pontrjagin} shows plots of the Twist density for different system sizes.

\section{Discussion} 

We now can draw several conclusions regarding the non-commutative and topological properties of the Moore-Read states with four anyons. Although the size of the systems we attempted are still small, the trend seen in the sequence of plots shown in Fig.~\ref{NonAbelianCurv} suggests that the Non-Abelian curvature is localized in the vicinity of the fixed anyons. This implies  a topological property for the Non-Abelian part of the mononodromies, i.e. their independence of the shape of braiding paths as long as the braiding path are far enough from the anyons.

Fig.~\ref{pontrjagin} demonstrates the existence of non-commutative monodromies. Indeed, if all the monodromies were commuting with each other, that will imply the existence of a gauge in which the fiber-bundle of the zero modes becomes trivial. But this will imply $\rho_{\text{\tiny{TW}}}$=0, which is contradicted by Fig.~\ref{pontrjagin}.

Fig.~\ref{pontrjagin} reveals much more, namely the splitting of the zero modes space as predicted by the fusion rules of the underlying CFT structure of the Moore-Read sequence.\cite{Feldman:2006ve,Bonderson:2006qf} For this, notice that the Twist density is practically zero when the mobile anyon comes near the fixed anyons and takes non-zero values only in between the fixed anyons. This implies that, although $|{\bf f}|$ takes appreciable values in the regions near the fixed anyons, the states are commutative. In other words, when the mobile anyon comes close to a fixed anyon, the two eigenmodes of the curvature matrix $\hat{F}(P)$ split the zero modes space into two sectors that don't mix with each other during the braiding. If things happen as predicted by the fusion rules of the underlying CFT structure of the Moore-Read sequence, then we should be able to read from the classic Bratelli diagram what these sectors are: they should be the q-spin $S$=0 and $S$=1 sectors.\cite{Bonderson:2006qf} For the sphere geometry, these sectors correspond to the even, respectively the odd numbers of electrons.

\begin{figure}
  \includegraphics[width=8.6cm]{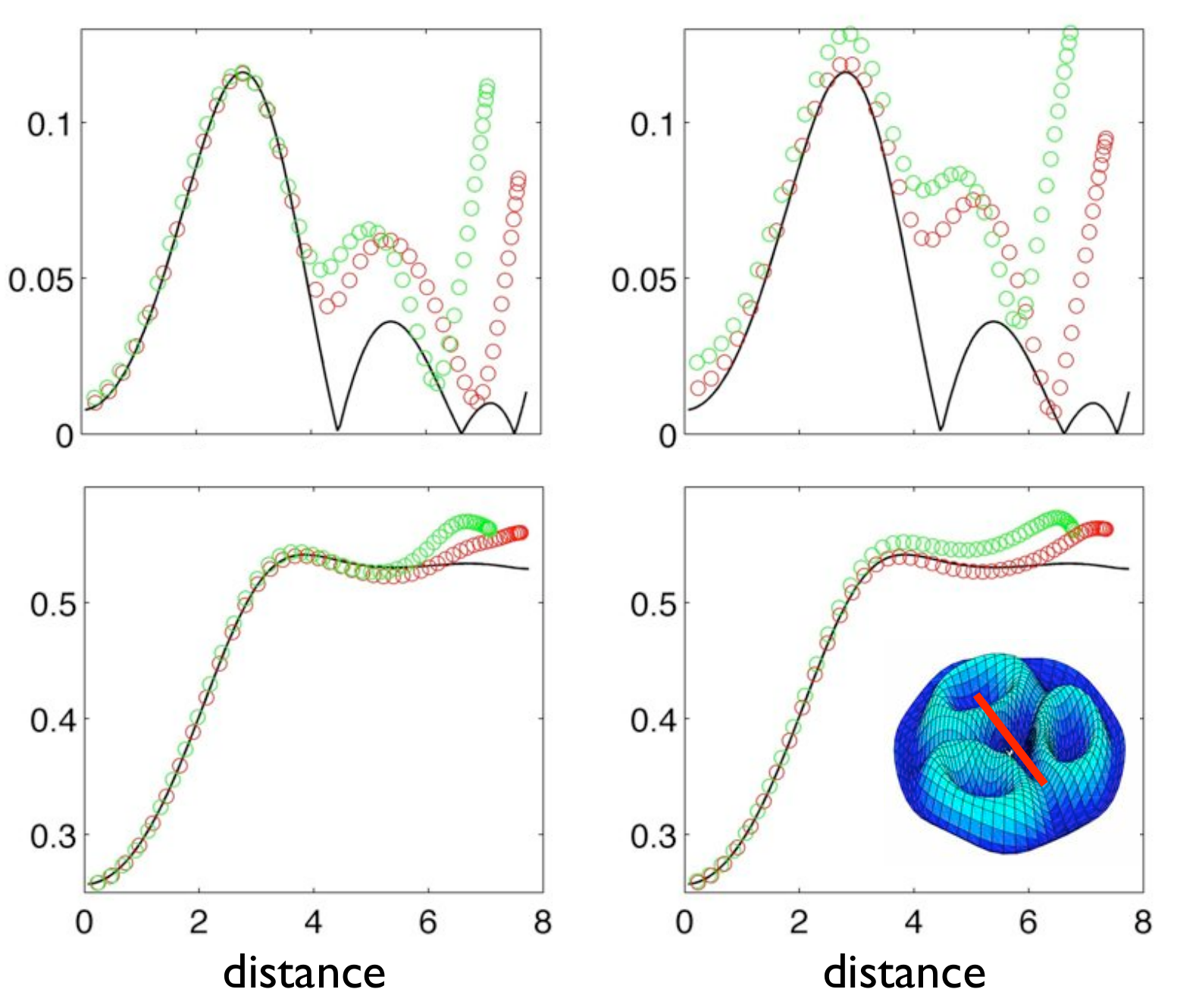}\\
  \caption{(Color online.) Comparison between the results obtained for the Abelian case (one flux added) and for the non-Abelian case (two flux added). The left/right panels refer to odd/even number of electrons. The solid lines in the upper/lower panels is the difference/sum between the Abelian Berry curvature for $N/N_\phi$=16/30 and $N/N_\phi$=15/29 (see Fig.~\ref{AbelianCurv}). The two open circle lines in the upper/lower panels show the values (multiplied by 2) of $f_0({\bf x})$/$|{\bf f}({\bf x})|$ recorded along the path shown in the inset. In the left panels, the dotted lines correspond to the system sizes $N/N_\phi$=15/29 and 13/25. In the left panels, the dotted lines correspond to $N/N_\phi$=14/27 and 12/23. }
  \label{AbelianVnonAbelian}
\end{figure}

To demonstrate that this is the case, we notice that $|{\bf f}(P)|$ plotted in Fig.~\ref{NonAbelianCurv} is equal to half the difference between the two eigenvalues of the curvature coefficient $\hat{F}(P)$ and that Tr$\hat{F}(P)$ plotted in Fig.~\ref{AbelianCurv} is equal to the sum of the two eigenvalues. Thus, near the fixed quasi-holes, the values of these quantities, properly normalized, should compare well with the values shown in the plots of the lower panels of Fig.~\ref{Curvature1Flux}. To verify that this is the case, we generated line-plots out the surface plots shown in Figs.~\ref{NonAbelianCurv} and \ref{AbelianCurv}. The line plots were generated by recording the points along the path shown in the inset of Fig.~\ref{AbelianVnonAbelian}. We used the data corresponding to the standard metric of the sphere. These line plots are compared in Fig.~\ref{AbelianVnonAbelian} with the absolute value of the data shown in Fig.~\ref{Curvature1Flux}. As one can see, the matching is almost perfect near the fixed anyon. By comparing two system sizes, we can also see the agreement becoming better as the size of the system is increased. We expect the agreement to become perfect as the sphere radius is taken to infinity. It is important to notice that there is agreement for both even and odd number of electrons cases.

Once we made this connection, we can we can go back and discuss the localization of the non-abelian curvature near the fixed quasiholes. The difference between the Berry phases for odd and even number of electrons, shown in Fig.~\ref{Phase1Flux}, converges extremely fast to $\pi$ [the plot suggests an exponential convergence]. From the connection made in the previous paragraph, this implies that the non-abelian curvature is also strongly [exponentially] localized near the fixed anyons. Unfortunately, the size of our systems is too small to allow a quantitative evaluation of this asymptotic decay behavior. Fig.~\ref{AbelianVnonAbelian} tells us how far are we from a converged situation, where the open circles should overlap with the continuous line.

We end this discussion Section by mentioning that we did compute several braiding monodromies and tried to compare their group properties under multiplication with the predictions following from the underlying CFT structure of the Moore-Read state derived in Ref.~\onlinecite{Nayak:1996mb}. Unfortunately, the size of our system is too small to see a strong correlation between the two. This can be seen from Fig.~\ref{AbelianVnonAbelian}, where the continuous line practically represent the thermodynamic limit of the non-abelian curvature and the open circles represent the same quantity for our largest finite system. The most problematic part is the large discrepancy seen at points in between the quasiholes (the end of the curves in Fig. ~\ref{AbelianVnonAbelian}), which affects any monodromy looping around a fixed anyon.

\section{Conclusions}

We have shown that the Hall sequences can be viewed as the zero modes of certain Hamiltonians, which were written down explicitly using many-particle creation operators. The compressible Hall states with the anyons fixed at definite locations were shown to be the zero modes of a pinning Hamiltonian, which was also written down explicitly. We have developed an efficient diagonalization algorithm for the pinning potential, which was subsequently used to explore some general properties and to map the quantum geometry of the Moore-Read states with two and four anyons.

Working with two anyons, we were able to give the first direct confirmation of the topological properties of the monodromies. The monodromy corresponding to a loop enclosing one of the anyons was found to differ by exactly a factor -1, when computed for even/odd number of electrons (which correspond to the q-spin $S$=0/1 sectors). This is the first explicit confirmation of the theoretical predictions based on the underlying CFT structure of the Moore-Read sequence.

By fusing the two anyons, we found that the electron density is different for the even/odd number of electrons. This is the first direct confirmation that one can determine if a quantum state is in the $S$=0/1 sectors by fusing the anyons and measuring the electron density. This type of measurement stands at the basis of read-in and read-out processes of TQC.\cite{Bonderson:2006qf}

Working with four anyons, we mapped the abelian and non-Abelian parts of the adiabatic curvature for the case of one itinerant anyon and three anyons fixed at the equator, in the most possible spread configuration. The mapping reveals that the non-Abelian curvature is strongly localized near the fixed quasiholes. We introduced the Twist density, which measures twisting of the zero modes during the adiabatic braiding. We found that the Twist density is practically zero near the fixed anyons, fact that signaled a splitting of the zero modes in two non-mixing sectors. Further analysis showed that this splitting is precisely the one implied by the fussion rules of the underlying CFT structure of the Moore-Read sequence.  

If the computations can be implemented to larger systems sizes, the present study open the possibility of: similar studies for higher level Hall sequences that support universal quantum computation; direct implementation and verification of the quantum gates found in Ref.~\onlinecite{Hormozi:2007ve} and a direct simulation of a quantum algorithm, including the read in, braiding and read out phases.

\noindent {\bf Acknowledgment} EP and DH both acknowledge support from the U. S. National Science Foundation (under MRSEC Grant No. DMR-0819860 and MRSEC grant/DMR-0213706) at the Princeton Center for ComplexMaterials. EP also wants to acknowledge support from the Research Corporation and from the office of the Provost of Yeshiva University.

\end{document}